\pdfoutput=1 \pdfsuppresswarningpagegroup=1 
\documentclass[
 aps,% 
 prb,%
 reprint, twocolumn,%
 groupedaddress,% 
 superscriptaddress,%
 amsfonts,%
 letterpaper,%
 footinbib,%
 noeprint%o
]{revtex4-2}

\RequirePackage{amsmath,amssymb,bm}
\RequirePackage{graphicx}
\RequirePackage[usenames,dvipsnames]{xcolor}

\usepackage{MnSymbol}

\usepackage{hyperref}
\hypersetup{
    colorlinks=true,
    citecolor=blue,
    linkcolor=magenta,
    filecolor=cyan,      
    urlcolor=magenta,
    pdfauthor = {Chuan Chen},
    pdfcreator = {\LaTeX\ and \flqq hyperref\frqq},
}

\usepackage[capitalise]{cleveref}

\graphicspath{ {./Figures/} }

\newcommand{\Fref}[1]{Fig.~\ref{#1}}
\newcommand{\Frefs}[1]{Figs.~\ref{#1}}
\newcommand{\Eqref}[1]{Eq.~\eqref{#1}}

\newcommand{\Secref}[1]{Section~\ref{#1}}

\newcommand{\Appref}[1]{Appendix~\ref{#1}}

\newcommand{\vk}{\vec{k}}
\newcommand{\vdelta}{\vec{\delta}}
\newcommand{\TaSe}{1T-TaSe\textsubscript{2}}
\newcommand{\TaS}{1T-TaS\textsubscript{2}}

\renewcommand{\Re}{\operatorname{Re}}
\renewcommand{\Im}{\operatorname{Im}}
\newcommand{\tapprox}{{\,\approx\,}}

\newcommand{\tequiv}{{\,\equiv\,}}
\newcommand{\teq}{{\,=\,}}

\newcommand{\tsim}{{\,\sim\,}}
\newcommand{\tlt}{{\,<\,}}
\newcommand{\tll}{{\,\ll\,}}
\newcommand{\tle}{{\,\le\,}}
\newcommand{\tlapprox}{{\,\lessapprox\,}}
\newcommand{\tgt}{{\,>\,}}
\newcommand{\tgg}{{\,\gg\,}}
\newcommand{\tge}{{\,\ge\,}}

\newcommand{\tplus}{{\,+\,}}

\newcommand{\meV}{\,\text{meV}}
\newcommand{\K}{\,\text{K}}

\DeclareMathOperator{\Tr}{Tr}

\begin{document}

\title{Competition of Spinon Fermi Surface and Heavy Fermi Liquids states from the
Periodic Anderson to the Hubbard model}

\author{Chuan~Chen}
\affiliation{Max-Planck Institute for the Physics of Complex Systems,
01187 Dresden, Germany}

\author{Inti~Sodemann}
\email[Correspondence to: ]{sodemann@pks.mpg.de}
\affiliation{Max-Planck Institute for the Physics of Complex Systems,
01187 Dresden, Germany}

\author{Patrick~A.~Lee}
\email[Correspondence to: ]{palee@mit.edu}
\affiliation{Department of Physics, Massachusetts Institute of Technology,
Cambridge Massachusetts 02139, USA}

\date{\today}

\begin{abstract}
We study a model of correlated electrons coupled by tunnelling to a layer of itinerant
metallic electrons, which allows to interpolate from a frustrated limit favorable to spin liquid
states to a Kondo-lattice limit favorable to interlayer coherent heavy metallic states.
We study the competition of the spinon fermi surface state and the interlayer coherent 
heavy Kondo metal that appears with increasing tunnelling.
Employing a slave rotor mean-field approach, we obtain a phase diagram and describe two
regimes where the spin liquid state is destroyed by weak interlayer tunnelling,
(i) the Kondo limit in which the correlated electrons can be viewed as localized spin moments
and (ii) near the Mott metal-insulator-transition where the spinon Fermi surface transitions
continuously into a Fermi liquid.
We study the shape of LDOS spectra of the putative spin liquid layer in the heavy Fermi liquid
phase and describe the temperature dependence of its width arising from
quasiparticle interactions and disorder effects throughout this phase diagram,
in an effort to understand recent STM experiments of the candidate spin liquid \TaSe \
residing on metallic 1H-TaSe\textsubscript{2}.
Comparison of the shape and temperature dependence of the theoretical and
experimental LDOS suggest that this system is either close to the localized Kondo limit, or in an
intermediate coupling regime where the Kondo coupling and the Heisenberg exchange
interaction are comparable.
\end{abstract}

\maketitle

\section{Introduction}
Since the pioneering proposal by Anderson \cite{Anderson1973,fazekas1974ground, Anderson1987}, there has been an
extensive quest to find quantum spin liquids (QSL) in materials
\cite{Broholm2020,Savary2016,Zhou2017}.
Recently, it has been suggested that certain layered
transition metal dichalcogenide compounds might harbour a
QSL state \cite{Law2017,He2018}.
In particular, \TaS, a material that undergoes a commensurate
charge density wave transition around $200 \K$ into a $\sqrt{13} \times \sqrt{13}$
star of David structure \cite{Rossnagel2011,Kratochvilova2017},
remains insulating to the lowest temperatures in spite of having an odd number of
electrons per star of David supercell, and yet shows no sign of any further conventional
ordering phase transition such as antiferromagnetism that would double the unit cell,
to the lowest measurable temperatures \cite{Klanjek2017}. A possible connection to Anderson's proposal of a spin liquid  was actually made from the very beginning \cite{tosatti1976nature}, but somehow forgotten.
The magnetic susceptibility of this compound remains nearly constant at low temperatures
\cite{Wilson1975} and
the material displays a finite linear in temperature specific heat coefficient \cite{Ribak2017}
indicative of a finite density of states at low energies.
Earlier experiments found no linear in temperature
heat conductivity \cite{Yu2017},
which was taken as evidence against itinerant carriers.
However, more recent experiments have shown a delicate 
sensitivity of heat transport to 
impurities~\cite{Murayama2020},
finding a finite linear in temperature heat conductivity
in the cleanest samples.
This indicates the presence of a finite density of states
of itinerant carriers, as expected for the spinon Fermi
surface state.
Moreover, band structure analysis \cite{Rossnagel2006} 
showed that a single narrow band crosses
the Fermi energy and is separated from other bands,
making it very likely that the low energy
electronic behaviour can be described by a single band 
Hubbard model.

A closely related compound, \TaSe, which also undergoes a commensurate charge density
wave transition into the star of David structure, is expected to display similar phenomenology.
While bulk \TaSe\ is metallic \cite{DiSalvo1974} , monolayer \TaSe\  was  studied by STM and
found to be a Mott insulator 
\cite{Chen2020}.
Recently Crommie and co-workers \cite{Wei2020} extended their study by placing a
monolayer of \TaSe\ on top of a 1H-TaSe\textsubscript{2} monolayer, which is metallic.
Surprinsingly their experiment has found that 
%although \TaSe\ is a Mott insulator
%with a gap near the Fermi energy, once it is placed on top of a monolayer metallic
%1H-TaSe\textsubscript{2} monolayer,
a Kondo-like resonance peak near the Fermi
energy develops in the tunnelling density of states.
It is important to emphasize that in these experiments the tunnelling tip is coupled
primarily to the originally insulating top layer of \TaSe. Therefore, taken at face value,
the appearance of a tunnelling density of states peak near zero bias may imply the destruction
of the presumed spin liquid that would exist for \TaSe\ in isolation and the formation of a
coherent metallic state by the coupling with the substrate metallic 1H-TaSe\textsubscript{2},
as it would be expected the classic problem of Kondo heavy metal formation.

These experimental findings motivate us to consider a model consisting of
a layer of correlated electrons coupled to a layer of non-interacting itinerant electrons via
tunnelling to study the competition of spinon Fermi surface states and the heavy Kondo metals.
There are two questions that we would like to address. First, the experimentalists found an 
excellent fit of the lineshape and its temperature dependence with that expected for the Kondo 
resonance of a single impurity Kondo problem \cite{Wei2020}. On the other hand, the actual 
system consists of a periodic array of local moments. Even if these are in the Kondo limit,
the low temperature state is expected to be a heavy Fermion metal. Would the formation of a
narrow coherent band lead to observable changes in the local density of states (LDOS)? Second,
how does the Heisenberg exchange coupling $J_H$ between the local moments compete with the
Kondo coupling $J_K$ that operates between the local moments and the conducting substrate?
This problem was considered by Doniach \cite{Doniach1977} for the case when the Heisenberg
coupling leads to an antiferromagnetic state. His conclusion is that the two relevant competing
energy scales are the Kondo temperature $T_K$ and the Heisenberg exchange scale $J_H$.
Note that at weak coupling $T_K $ is exponential small in terms of the Kondo coupling $J_K$.
This would suggest that a very weak $J_H$ is sufficient to destroy the Kondo effect.
If the experiment was interpreted as being in the Kondo limit, this places a rather small upper
bound on $J_H$ of about 50K, since the scale $T_K$ is estimated to be about 50K from the 
experimental fit \cite{Wei2020}. 
With such a small Heisenberg coupling, the interpretation of the monolayer \TaSe\ as a
spin liquid 
is brought into question. We note that the situation may change when the coupling becomes 
strong, and it may also change in frustrated spin models where the spin liquid state may be 
favored over the anti-ferromagnet. Notice that in the resonating valence bond (RVB) picture, the 
quantum spin liquid is viewed as the superposition of singlet formed between local moment pairs, 
while the Kondo phenomenon arises from the singlet formation between the local moment
and the conduction electron spin.
The competition between different ways of forming singlets may well be different 
from the competition with an anti-ferromagnet considered by Doniach. 
With this in mind, we will consider a model that is suffiicently general to include the Hubbard 
interaction ($U$) for the correlated electrons that reside in the putative spin liquid layer,
which hop with an amplitude ($t_d$) within this layer,
and a tunnelling amplitude ($V$) to the itinerant electrons residing in the putative metallic layer,
which hop with an amplitude ($t_c$) within their own layer, as dipicted in
\Fref{fig:schematic}.
This model therefore interpolates naturally between the periodic Anderson model
($t_d \rightarrow 0$) where it would capture the physics of the formation of the
interlayer coherent heavy Kondo metal \cite{Hewson1993,Coleman2015}
and the pure Hubbard limit ($V \rightarrow 0$) where it would capture the traditional
scenario for the appearance of the spinon Fermi surface state near the Mott transition
\cite{Florens2004,Lee2005,Motrunich2005}.
We note in passing that this model has been recently employed to understand
ARPES spectra in PdCrO\textsubscript{2} \cite{Sunko2020}, however, in this material the insulating layers are believed to be strong Mott insulators with $120^{\circ}$ spin anti-ferromagnetic order.

One of the central quantities of our interest will be the 
LDOS of the
putative spin liquid layer, which is what has been measured in the aforementioned STM
experiments. We are particularly interested in understanding the temperature dependence
of the width of the LDOS peak, which can be used to try to learn about the microscopic
parameters of the putative spin liquid and its coupling to the metal, and can guide us in
determining where the system is
likely to lie in the parameter space of our Hubbard-Anderson periodic model.
Although an unambiguous quantitative description of the temperature
dependence is challenging because it is controlled by the interplay of intrinsic quasi-particle
lifetimes and extrinsic effects such as disorder induced broadening, we believe that
our modelling is consistent with the system to be either close to the periodic Anderson model
limit or in an intermediate coupling regime where
the Kondo coupling and the Heisenberg exchange interaction are comparable,
as we will discuss in detail.
In the latter case, we cannot extract a tight bound on $J_H$ based on the experimental data.

Our paper is organized as follows: \Secref{sec:sr-mean-field} sets up the model and describes
the mean-field slave rotor approach that we employ to tackle it.
\Secref{sec:mean-field-pd} presents the solution of this mean field under a wide range of
parameters, including not only the interplay between spinon Fermi surface and heavy metal
but also the possibility of competing with Kondo insulating states.
\Secref{sec:ldos} is devoted to a detailed analysis of the LDOS spectra and temperature 
dependence of the LDOS width and the comparison with STM experiments.
\Secref{sec:discuss} summarizes and further discusses our main findings.
We have relegated some of the technical details of the mean-field treatment to 
\Appref{sec:append-rotor}.
In \Appref{sec:append-ldos} 
we revisit the classic result of the temperature dependence
of the single impurity Anderson model and %points out a correction to the
give a more thorough derivation of the width of the
Kondo resonance.
%that has been missed in previous interpretations of experiments

%(compare \Eqref{eq:Gamma exact} and \Eqref{eq:width-exp}).

\section{Model and Slave Rotor approach}\label{sec:sr-mean-field}
\begin{figure}
\centering
\includegraphics[width=0.46\textwidth]{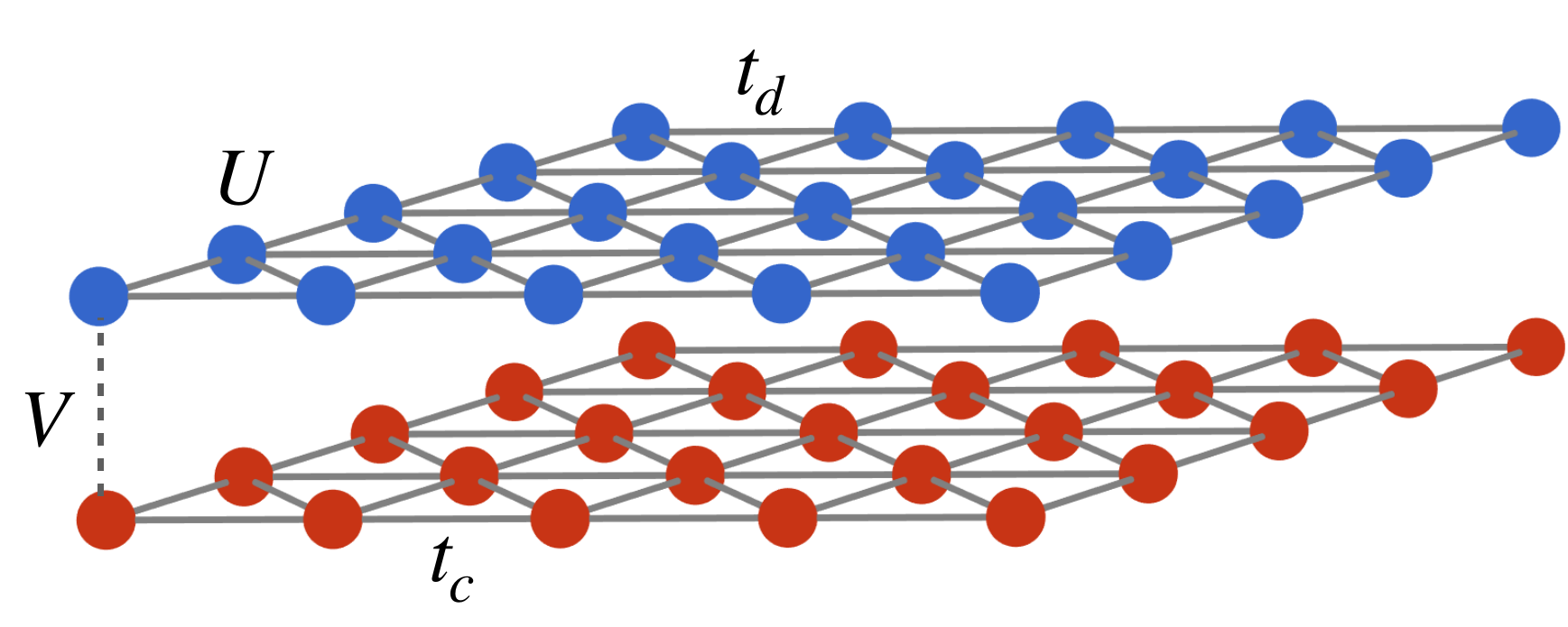}
\caption{
Schematic of the model.
The electrons in the top layer (blue) are correlated, with nearest neighbour hopping $t_d$
and an on-site Hubbard interaction $U$.
The bottom layer (red) hosts itinerant electrons with nearest neighbour hopping $t_c$.
There is also an inter-layer tunneling $V$.
}
\label{fig:schematic}
\end{figure}

We consider a model of two-species of fermions residing in a triangular lattice that
interpolates naturally between the Hubbard model and the periodic Anderson model.
The microscopic Hamiltonian of the system has the form:
\begin{align} \label{eq:H}
H = & -t_d \sum_{\langle i,j \rangle,\sigma} d_{i,\sigma}^{\dagger} d_{j,\sigma} 
+ \sum_{i} n_{d,i} ( \epsilon_d^{(0)} - \mu_F ) \nonumber \\
&  - t_c \sum_{ \langle i,j \rangle,\sigma} c_{i,\sigma}^{\dagger} c_{j,\sigma} 
+ \sum_{i} n_{c,i} ( \epsilon_c^{(0)} - \mu_F ) \nonumber \\
& + \frac{U}{2} \sum_i (n_{d,i} - 1)^2
- V \sum_{i,\sigma} \left( c_{i,\sigma}^{\dagger} d_{i,\sigma} + h.c. \right).
\end{align}
Here the electrons created by $c^{\dagger}$ are viewed as the ``itinerant'', and those
created by $d^{\dagger}$ as the correlated ones.
A schematic of the system is shown in \Fref{fig:schematic}.
In the limit in which the correlated electrons are
localized, $t_d=0$, this model reduces to the Periodic Anderson model, and in the limit in which
the two specifies are decoupled, $V=0$, the Hamiltonian for the correlated electrons reduces
to the Hubbard model. 
We would like to employ a formalism capable of handling the various regimes of this model,
and in particular the single occupancy constraints that appear in the large $U$ limit.
For this purpose we resort to the slave rotor mean-field approach.
According to the slave rotor method \cite{Florens2004,Zhao2007}, the $d$-electron can be
represented by a bosonic rotor, $\theta_i$, and a fermionic spinon $f_{i,\sigma}$
degrees of freedom:
$d_{i,\sigma} \tequiv e^{i \theta_i}f_{i,\sigma}$, 
with the constrain $n_{\theta,i} \tplus n_{f,i} \teq 1$.
The Hamiltonian can be then written in terms of these partons as follows:
\begin{align} \label{eq:H_sr}
H = & -t_d \sum_{\langle i,j \rangle,\sigma} e^{-i\theta_i}e^{i\theta_j} f_{i,\sigma}^{\dagger}
f_{j,\sigma} 
+ \sum_{i} n_{f,i} ( \epsilon_d^{(0)} - \mu_F ) \nonumber \\
&  - t_c \sum_{ \langle i,j \rangle,\sigma} c_{i,\sigma}^{\dagger} c_{j,\sigma} 
+ \sum_{i} n_{c,i} ( \epsilon_c^{(0)} - \mu_F ) \nonumber \\
& + \frac{U}{2} \sum_i n_{\theta,i}^2
- V \sum_{i,\sigma} \left( e^{i\theta_i} c_{i,\sigma}^{\dagger} f_{i,\sigma} + h.c. \right).
\end{align}
% ------------------------------------------------------------------------------
% FIGURE
% ------------------------------------------------------------------------------
% ------------------------------------------------------------------------------
%
\subsection{Mean-field theory}
In the spirit of a mean-field theory we approximate the ground state of
\Eqref{eq:H_sr} by a direct product of a rotor state and a spinon state. The constrain on
the rotor and spinon occupation is satisfied on average:
\begin{equation}\label{eq:constrain}
\langle n_{\theta,i} \rangle \tplus \langle n_{f,i} \rangle \teq 1.
\end{equation}
Since the rotor and spinon degrees of freedom are assumed to be disentangled,
we write the mean-field Hamiltonian as the sum of a rotor part and a fermionic
part, i.e.,  $H_{\text{mf}} \teq H_f + H_\theta$, with
\begin{subequations}
\begin{align}
H_{f} &= -T_f \sum_{\langle i,j \rangle,\sigma} f_{i,\sigma}^{\dagger} f_{j,\sigma} 
+ \sum_{i} n_{f,i} ( \epsilon_d^{(0)} + \lambda - \mu_F )  \label{eq:H_f-ini} \nonumber \\
& \ - t_c \sum_{ \langle i,j \rangle,\sigma} c_{i,\sigma}^{\dagger} c_{j,\sigma} 
+ \sum_{i} n_{c,i} ( \epsilon_c^{(0)} - \mu_F ) \nonumber \\
& \ - V_f \sum_{i,\sigma} c_{i,\sigma}^{\dagger} f_{i,\sigma} + h.c., \\
H_{\theta} &= -2 \sum_{\langle i, j \rangle} T_{\theta} e^{-i\theta_i}e^{i\theta_j} + \sum_i \frac{U}{2} n_{\theta,i}^2
+ \lambda n_{\theta,i} - 4 V_{\theta} \cos(\theta_i),
\label{eq:H_theta} \\
T_f &= t_d \langle e^{-i\theta_i} e^{i\theta_j} \rangle_{\theta}, \\
V_f &= V \langle e^{i\theta_i} \rangle_{\theta}, \label{eq:Vf} \\
T_{\theta} &= t_d \langle f_{i,\sigma}^{\dagger} f_{j,\sigma} \rangle_f, \\
V_{\theta} &= V \langle c_{i,\sigma}^{\dagger} f_{i,\sigma} \rangle_f,
\end{align}
\end{subequations}
here a Lagrange multiplier $\lambda$ is introduced to maintain the constrain
\Eqref{eq:constrain}.
The quasiparticle residue of correlated $d$ electron is
$\langle e^{i\theta_i} \rangle \tequiv \Phi$. This can be regarded as the order parameter for the
metallic phase: when it is non-zero there will be a coherent tunnelling between the
spinon and itinerant electrons.
In this work, we will concentrate on the competition of this correlated metallic state and a more
exotic state, known as the spinon Fermi surface state, that arises when $\Phi=0$ and the
spinon, $f$, has a Fermi surface.

We expect that the essence of the competition between these phases does not depend substantially on the details of the fermion dispersions, and therefore, in order to simplify analytical treatment, we will approximate the band structure for spinons ($f$) and
itinerant electrons ($c$) by simple parabolic bands:
\begin{subequations}
\begin{align}
H_f &= \sum_{k,\sigma} f_{k,\sigma}^{\dagger} f_{k,\sigma} \epsilon_{f,k}
+ c_{k,\sigma}^{\dagger} c_{k,\sigma} \epsilon_{c,k}
- V_f \left( c_{k,\sigma}^{\dagger} f_{k,\sigma} + h.c. \right),
\label{eq:H_f} \\
\epsilon_{f,k} &= \frac{3}{2} T_f \left( k^2 - \frac{\Lambda^2}{2} \right) + \lambda - \mu_F,
\label{eq:spinon-bands} \\
\epsilon_{c,k} &= \frac{3}{2} t_c \left( k^2 - \xi \frac{\Lambda^2}{2} \right) - \mu_F,
\label{eq:c-bands}
\end{align}
\end{subequations}
here $\Lambda$ is a cut-off on $k$-space  intended to mimic the finite size of the
Brillouin zone which can be determined by equalling
$\pi \Lambda^2$ to the area of triangular lattice's Brillouin zone, the lattice constant $a_0$ is taken to be $1$.
The dimensionless parameter
$\xi$ in $\epsilon_{c,k}$ reflects the occupancy of $c$ electrons when $c$ and $f$
fermions are decoupled (since in such case $\lambda \teq 0$ and $\mu_F \teq 0$,
see discussions in the following section):
the number of $c$ electron per site is $\xi$ when the dispersion
$\epsilon_{c,k}$ is particle ($t_c \tgt 0$), and $2-\xi$ with
hole like dispersion ($t_c \tlt 0$). See \Fref{fig:bands} for a schematic illustration.
\subsection{Expectation values of the rotor operators}
Notice that even after the mean field decoupling, the rotor Hamiltonian $H_{\theta}$ is still
essentially a $2D$ quantum XY model with a transverse field which is not amenable to analytic
treatment. 
Therefore, one has to make further approximations.

We are interested in solutions that respect time-reversal and translational symmetry and
that have no flux per unit cell.
Therefore we seek for self-consistent solutions where $\Phi$ is uniform and real.
To do so, we perform an additional self-consistent mean-field treatment of $H_\theta$
by introducing an effective single-site rotor Hamiltonian:
\begin{subequations}
\begin{align}
H_{\theta}^{(1)} &= -K_\theta \left( e^{i\theta} + e^{ -i\theta} \right)
+ \frac{U}{2} n_{\theta}^2 + \lambda n_{\theta}, \\
K_{\theta} &= 2 z T_\theta \Phi +2 V_\theta, \label{eq:K_theta-def}
\end{align}
\end{subequations}
with $z$ being the lattice coordination ($z \teq 6$ for triangular lattice).
To lowest order in perturbation theory in $K_\theta/U$ ($\lambda \teq 0$
since we are interested in half-filled spinon and the constrain \Eqref{eq:constrain} leads to
$\langle n_{\theta,i} \rangle \teq 0$) we have
$\Phi = 4 K_\theta/U$.
On the other hand, in the opposite limit in which $K_\theta/U \tgg 1$, we have
$\theta \tapprox 0$ and thus $\Phi \teq \langle e^{i\theta} \rangle \teq 1$.
Moreover, since $\Phi \teq \langle e^{i\theta} \rangle$ is never greater than one,
we introduce the following natural interpolation between these limits:
\begin{equation}
\Phi = \frac{K_\theta}{\sqrt{(U/4)^2+K_\theta^2}},
\end{equation}
or equivalently,
\begin{equation} \label{eq:K_theta}
K_\theta = \frac{U}{4} \frac{\langle e^{i\theta} \rangle}{\sqrt{1-\langle e^{i\theta} \rangle^2}},
\end{equation}

Although the above mean field treatment captures well the behavior of the residue $\Phi$,
it ignores completely the nearest neighbour rotor correlations, which are essential in order
to obtain a dispersion for the spinon.
To capture these, and since $V_\theta$ is small near the metal to insulator phase transition,
we will approximate their value by performing a perturbative calculation directly with the more
complete rotor Hamiltonian $H_\theta$ from \Eqref{eq:H_theta},
which contains the $U$ and $T_\theta$ terms only,
\begin{equation} \label{eq:H_r}
\tilde{H}_\theta = \frac{U}{2} \sum_{i} n_{\theta,i}^2
- 2 T_\theta \sum_{\langle i,j \rangle}e^{-i\theta_i}e^{i\theta_j},
\end{equation}
which leads to the following nearest neighbor rotor correlations:
\begin{equation} \label{eq:rotor-correlation}
\langle e^{-i\theta_i}e^{i\theta_j} \rangle \approx \frac{4 T_\theta}{U},
\end{equation}
it should be noted that these nearest-neighbor rotor correlations from
\Eqref{eq:rotor-correlation} are needed to reproduce the spinon
bandwidth which is expected to be given by the Heisenberg exchange coupling scale
$J_H=4t_d^2/U$. The expressions above are all zero temperature results.
The finite temperature version of these formulae are discussed in \Appref{sec:append-rotor}.
\subsection{Expectation values of the fermion operators}
The fermionic mean-field Hamiltonian is free from interactions and can be diagonalized exactly.
Because we are already accounting for spinon hopping in the spin liquid phase at $V \teq 0$,
the correlator $\langle f_{i,\sigma}^{\dagger} f_{j,\sigma} \rangle$ is not expected to change
much during the spin-liquid to heavy-metal phase transition,
so we will simply approximate its value when $c$
and $f$ fermions are decoupled from each other ($V_f = 0$ in the insulating phase):
\begin{equation} \label{eq:chi_0}
\langle f_{i,\sigma}^{\dagger} f_{j,\sigma} \rangle = \frac{1}{N}\sum_k e^{i \vk \cdot \vdelta}
n_F(\epsilon_{f,k}) \equiv \chi_0,
\end{equation}
with $n_F$ being the Fermi-Dirac distribution function: $n_F(x) = 1/\left( e^{\beta x}+1 \right)$, $\delta$ is the distance between sites $i$ and $j$, and $N$ is the total number of lattice sites in \Eqref{eq:chi_0}.
thus $T_\theta = t_d \chi_0$.
As for the hybridization between the itinerant electrons
and spinons, one obtains:
\begin{subequations}
\begin{align}
& \langle c_{i,\sigma}^{\dagger} f_{i,\sigma} \rangle = V_f \chi_{cf}, \label{eq:cf} \\
& \chi_{cf} = -\frac{1}{2N} \sum_k \frac{n_F(E_{1,k}) - n_F(E_{2,k})}{
\sqrt{ (\frac{\epsilon_{f,k} - \epsilon_{c,k}}{2})^2+V_f^2} }. \label{eq:chi_cf}
\end{align}
\end{subequations}
It should be noted that \Eqref{eq:cf} is an exact result of solving the free fermionic Hamiltonian 
$H_f$, although in the $V_f \rightarrow 0$ limit, the $\chi_{cf}$ reduces to the
$c$-$f$ hybridization susceptibility of the $c$-$f$ decoupled Hamiltonian.
The quasi-particle energy dispersions read (see \Fref{fig:bands}):
\begin{equation} \label{eq:E_12}
E_{1/2,k} = \frac{\epsilon_{f,k} + \epsilon_{c,k}}{2} \pm
\sqrt{ \left( \frac{\epsilon_{f,k} - \epsilon_{c,k}}{2} \right)^2 + V_f^2 },
\end{equation}
and the occupancy of spinon reads:
\begin{subequations}
\begin{align} \label{eq:n_f}
\langle f_{i,\sigma}^\dagger f_{i,\sigma} \rangle &= \frac{1}{N} \sum_k
\cos^2(\alpha_k) n_F(E_{1,k}) + \sin^2(\alpha_k) n_F(E_{2,k}), \\
\cos(2\alpha_k) &= \frac{\epsilon_{f,k} - \epsilon_{c,k}}{2}/
\sqrt{ \left( \frac{\epsilon_{f,k} - \epsilon_{c,k}}{2} \right)^2 + V_f^2 },
\end{align}
\end{subequations}
\subsection{Self-consistent equations}
Once the expressions for the expectation values of the rotor and fermions are obtained,
the self-consistent equations for the order parameter $\Phi$ can be derived,
from \cref{eq:K_theta-def,eq:K_theta,eq:cf},
one can show that:
\begin{equation} \label{eq:self-cons-full}
\frac{\Phi}{8} \left( \frac{1}{\sqrt{1-\Phi^2}}
- 8 z \frac{t_d}{U} \chi_0 \right)
= \Phi \frac{V^2}{U} \chi_{cf}.
\end{equation}
Therefore, one needs to solve \Eqref{eq:self-cons-full} along with the constrain \Eqref{eq:constrain} and
$\langle n_{f,i} \rangle \teq 1$.
\Eqref{eq:self-cons-full} always has a trivial solution
$\Phi \teq \langle e^{i\theta_i} \rangle \teq 0$, and the non-trivial solution of
$\langle e^{i\theta_i} \rangle$ satisfies:
\begin{equation} \label{eq:self-cons}
\frac{1}{8} \left( \frac{1}{\sqrt{1 - \Phi^2}}
- 8 z \frac{t_d}{U} \chi_0 \right)
= \frac{V^2}{U} \chi_{cf}.
\end{equation}
It should be noted that the ``susceptibility'' $\chi_{cf}$ also depends on 
$\Phi$, through its dependence on $V_f$ in \Eqref{eq:chi_cf}, which in turn
depends on $\Phi$ via \Eqref{eq:Vf}.
\section{Mean-field phase diagram and mean-field properties.}\label{sec:mean-field-pd}
\begin{figure}
\centering
\includegraphics[width=0.45\textwidth]{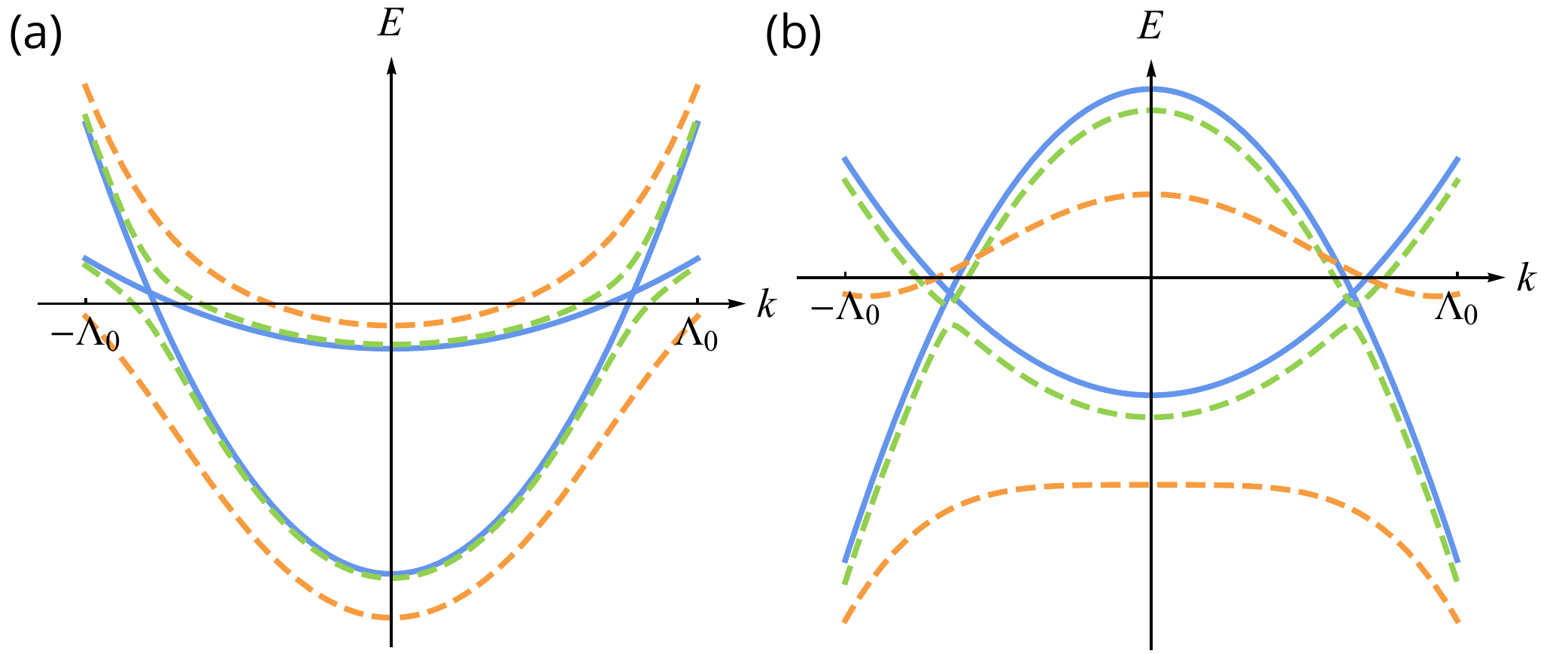}
\caption{
Schematic of the band dispersion.
(a) Particle-particle dispersion (with $\xi \tgt 1$).
Blue solid lines indicate the $\epsilon_{f,k}$ and $\epsilon_{c,k}$ in the
spin liquid phase;
green dashed lines stand for the $E_{1,k}$ and $E_{2,k}$ for small $V_f$, where both bands
cross the Fermi level and there are two Fermi surfaces;
the orange dashed lines are when $V_f$ is large such, so that the $E_{2,k}$ band
is fully occupied and $E_{1,k}$ is partly occupied to maintain
the half filling of the spinon.
(b) Particle-hole dispersion ($\xi \tlt 1$).
For small $V_f$ (green dashed line) only $E_{1,k}$ crosses the Fermi level and has two Fermi
surfaces while the $E_{2,k}$ is fully occupied; when $V_f$ is large enough (orange dashed lines) there is only one Fermi surface. 
}
\label{fig:bands}
\end{figure}

To explore the phase transition between the spin liquid and heavy metal phases, it is important
to distinguish the cases with the band dispersions of the $d$-electron and itinerant electrons
being particle-particle like ($t_d \tgt 0$ and $t_c \tgt 0$) and particle-hole like
($t_d \tgt 0$ and $t_c\tlt 0$).
Here we discuss in detail the behavior when the itinerant fermion has higher density
(larger Fermi surface area) than the spinon, which is most relevant to the recent experiments
\TaS\ and \TaSe.
Namely we will take the paramter $\xi$, that controls the density of the itinerant electrons in \Eqref{eq:c-bands}, to have a range of $1 \tle \xi \tlt 2$ for the particle-particle case
and $0 \tle \xi \tlt 1$ for the particle-hole case (this leads to
$n_c \tge n_f$ in the insulating phase), see \Fref{fig:bands} for an illustration.
% ------------------------------------------------------------------------------
% FIGURE
% ------------------------------------------------------------------------------
% ------------------------------------------------------------------------------
%
\subsection{Particle-particle dispersion}
In this section we discuss the situation for particle-particle like dispersions.
As mentioned before, there are two competing phases in our phase diagram:
the spin liquid phase and the heavy metal phase (see \Fref{fig:phase-diagram-pp} for an example
of the phase diagram). The phases are determined by whether order parameter
$\Phi$ is finite (heavy metal) or zero (spin liquid). When $t_d \tsim 0$, the model reduces to
a periodic Anderson model and the transition from spin liquid to heavy metal is of the form
of a weak coupling instability. On the other hand, for larger $t_d/U \tsim 1/8$ and $V \teq 0$,
the system exhibits a metal-insulator (Mott) transition, as one expects from a
Hubbard model.
The goal of next section is to determine how the phase boundary evolves between these two regimes.
\subsubsection{Phase boundary}
\begin{figure}
\centering
\includegraphics[width=0.49\textwidth]{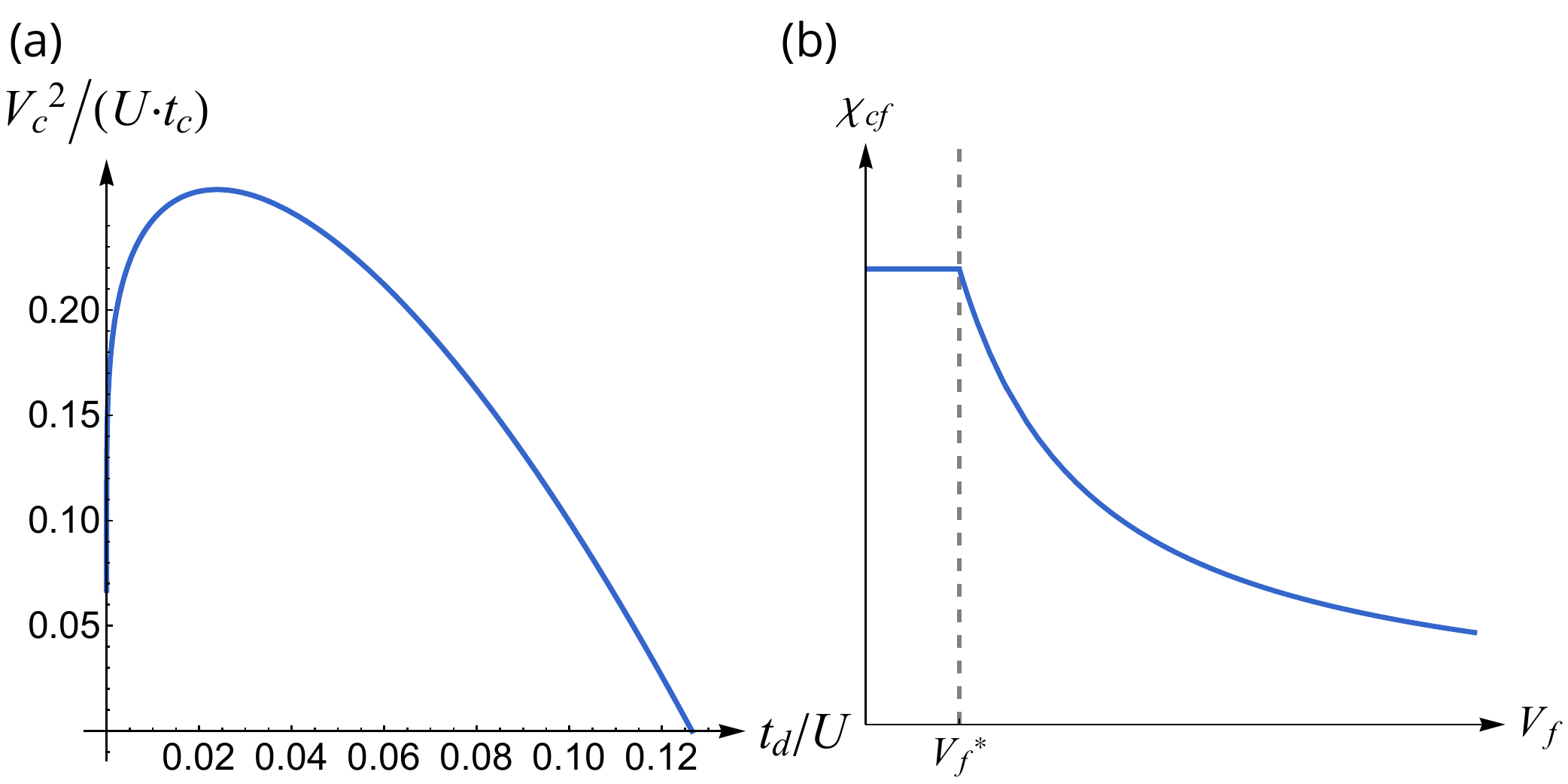}
\caption{
(a) Phase boundary between spin liquid (below blue curve) and heavy metal
with particle-particle dispersion and $\xi \teq 1.2$.
As $t_d \rightarrow 0$, the critical coupling $V_c^2/U$ is suppressed logarithmically with $t_d/U$;
when $V \teq 0$ (horizontal axis), the metal-insulator transition occurs at $t_d/U \sim 1/8$.
Near this critical point, the $V_c^2/U$ has a linear dependence on $t_d/U$.
(b) Plot of the $\chi_{cf}$ with $T_f \teq 0.1 t_c$. $\chi_{cf}$ saturates at small $V_f$,
while for $V_f \tgt V_f^*$, it is a decreasing function of $V_f$.
}
\label{fig:pb-chi-pp}
\end{figure}

The phase boundary is obtained when $\Phi \teq 0$ is a solution
of \Eqref{eq:self-cons}.
According to the constraint from \Eqref{eq:constrain} and $\langle n_{f,i} \rangle \teq 1$,
we have that $\langle n_{\theta,i} \rangle \teq 0$.
This leads to a value $\lambda=0$ for the Lagrange multiplier in \Eqref{eq:H_theta}.
Thus one just needs to self-consistently adjust the chemical potential $\mu_F$ such that
the spinon is half-filled.
Along the phase boundary, since $c$ and $f$ fermions are decoupled, 
this can be satisfied by setting $\mu_F \teq 0$, which leads to
$n_{f,i} \teq 1$ and $n_{c,i} \teq \xi$, which corresponds to two Fermi surfaces from the two
bands with Fermi momentum $k_{F,f} \teq \Lambda/\sqrt{2}$ and
$ k_{F,c} \teq \Lambda \sqrt{\xi/2}$.
In this case the susceptibility of $c$-$f$ coupling from \Eqref{eq:chi_cf},
reduces to:
\begin{align}
\chi_{cf}^{(0)} &= -\frac{1}{N} \sum_k
\frac{n_F(\epsilon_{f,k}) - n_F(\epsilon_{c,k})}{\epsilon_{f,k} - \epsilon_{c,k}} \nonumber \\
&= \frac{1}{\Lambda^2}\frac{2}{3} \frac{1}{T_f - t_c} \ln \left( \frac{T_f}{t_c} \right).
\end{align}
It is interesting to notice that the $\chi_{cf}^{(0)}$ is independent of $\xi$;
in other words, the density of itinerant electrons.
This implies that the phase boundary is insensitive to the
$c$ electron's density within the parabolic band approximation.
The critical value at which the residue $\Phi$ and the hibridization between the itinerant and
correlated electron, $V_f$, become simultaneously non-zero is given by:
\begin{equation} \label{eq:V_c}
\frac{V_c^2}{U t_c} = \frac{1}{8}\left( 1-8 z \frac{t_d}{U}\chi_0 \right) 
(\frac{4 t_d^2 \chi_0}{U t_c} - 1) \frac{\frac{3}{2} \Lambda^2}{\ln\left( \frac{4 t_d^2 \chi_0}{t_c U} \right)}.
\end{equation}
A plot of the phase boundary in this case can be found in \Fref{fig:pb-chi-pp}(a).
As it approaches the Anderson ($t_d \rightarrow 0$) limit, the critical $V_c^2/U$ has a
logarithmic dependence on $t_d/U$.
This means that in the local moment limit, the heavy Fermi liquid phase is destabilized
by a weak Heisenberg coupling, $J_H \tsim t_d^2/U$, comparable to the Kondo scale,
$T_K \tsim \rho^{-1} e^{-1/J_K \rho}$ (with $J_K \tsim V^2/U$ and $\rho^{-1} \tsim t_c$).
This is responsible for the sharp narrowing of the region of the Heavy Fermi liquid phase
in the local moment limit $V^2 \tll t_c U$, and $t_d \tll U$, as shown in \Fref{fig:pb-chi-pp}(a).
Around the axis $V \teq 0$ we recover the physics of the spin-liquid to metal
(Mott transition) in the conventional Hubbard model
with the spin-liquid to metal transition (see Ref.~\cite{Florens2004}) occurring at $t_d/U=1/(8 z\chi_0)$, which in the case of the triangular lattice corresponds to $t_d/U \tsim 1/8$ and is in line with previous cluster mean-field
calculation \cite{Zhao2007}.
% ------------------------------------------------------------------------------
% FIGURE
% ------------------------------------------------------------------------------
% ------------------------------------------------------------------------------
%
\subsubsection{Turning on of the heavy fermion phase}
As one enters the heavy fermion metallic phase ($\Phi$ becomes finite),
both the $E_{1,k}$ and $E_{2,k}$ bands cross the
Fermi level (as indicated by the green dashed lines in \Fref{fig:bands}(a)).
According to \Eqref{eq:n_f}, the spinon density in this case is:
\begin{align}
\langle f_{i,\sigma}^\dagger f_{i,\sigma} \rangle &= \frac{k_{F1}^2+k_{F2}^2}{2\Lambda^2}
+ \frac{\sum_{\alpha = c,f} \epsilon_{\alpha,k_{F1}}
+ \epsilon_{\alpha,k_{F2}}}{3\Lambda^2 (t_c - T_f)},
\end{align}
by requiring this to be $1/2$, one can obtain $\mu_F \teq 0$ (with $\lambda \teq 0$).
It can be shown that in this case, the susceptibility is simply a constant:
\begin{equation} \label{eq:chi_cf-cons}
\chi_{cf} = \frac{2}{3} \frac{1}{t_c \Lambda^2} \frac{1}{\frac{4t_d^2 \chi_0}{U t_c} - 1}\ln(\frac{4 t_d^2 \chi_0}{U t_c}).
\end{equation}
Notice that $\chi_{cf}$ is independent of $V_f$ (or $\Phi$), which is a consequence of the
parabolic model. Physically $\chi_{cf}$ should be a monotonically decreasing
function of $V_f$ for a general band dispersion, but we conclude from the above that it is
weakly dependent on these parameters whenever the bands can be approximated by parabolas.
Nevertheless, \Eqref{eq:chi_cf-cons} still unveils an important effect of the correlated fermion
hopping $t_d$,
which is to set a ``cut-off'' to $\chi_{cf}$,
as depicted in \Fref{fig:pb-chi-pp}(b). Such cut-off would otherwise be absent in the pure periodic
Anderson model ($t_d \rightarrow 0$) and we would have that
$\chi_{cf} \rightarrow \infty$ as $V_f \rightarrow 0$.
This divergence is responsible for the
weak-coupling (Kondo) instability of the periodic Anderson model that leads
to the formation of the heavy Fermi liquid state.

On the other hand, there is a further phase transition that appears within the heavy Fermi liquid
state, associated with the disappearance of one of the Fermi surfaces while preserving the net
Luttinger volume, at large $V_f$.
This occurs when $V_f$ is larger than some critical value
$V_f^* \teq \frac{3}{2} \frac{\Lambda^2}{2} \sqrt{T_f t_c ( 2-\xi )}$, for which we have that
$E_{2,\Lambda} \tlt 0$, so the $E_{2,k}$ band is fully occupied and there is only one Fermi surface
associated with the band $E_{1,k}$ (see yellow dashed lines in \Fref{fig:bands}(a)).
In this case, the density of spinon reads:
\begin{widetext}
\begin{equation}
\langle f_{i,\sigma}^\dagger f_{i,\sigma} \rangle = \frac{k_{F1}^2 + \Lambda^2}{2\Lambda^2}
+ \frac{ \epsilon_{f,k_{F1}} + \epsilon_{c,k_{F1}}
+ \sqrt{ \left( \epsilon_{f,\Lambda} - \epsilon_{c,\Lambda}\right)^2 + 4V_f^2 }}
{3\Lambda^2 (t_c-T_f)},
\end{equation}
\end{widetext}
and the $\mu_F$ can be determined by requiring
$\langle f_{i,\sigma}^\dagger f_{i,\sigma} \rangle \teq 1/2$.
In this case the susceptibility $\chi_{cf}$ is no longer independent of $V_f$
(we do not show the explicit expression here since it is too lengthy).
\Fref{fig:pb-chi-pp}(b) shows a plot the
$\chi_{cf}$ as a function of $V_f$ for a specific parameterization.
As mentioned before, a finite $t_d$ sets a ``cut-off'' to the $\chi_{cf}$, moreover,
the critical $V_f^*$ will also decrease as $t_d$ decreases.
This role of $t_d$ as a cutoff of the $\chi_{cf}$ susceptibility leads to an increasing value of the critical $V$ as $t_d$ increases at extremely small
values of $t_d$, as shown in \Fref{fig:pb-chi-pp}(a).
In other words, the larger the value of $t_d$ the smaller the susceptibility to induce the mixing
between the itinerant and correlated fermions.

However, the physical role of $t_d$ is not exclusively to cutoff $\chi_{cf}$.
It is clear from the \Fref{fig:pb-chi-pp}(a) that at sufficiently large $t_d$
the critical $V$ starts to decrease as $t_d$ increases.
The other physical role of $t_d$ can be understood
from the self-consistent equation for the residue $\Phi$, \Eqref{eq:self-cons},
where we see that the hopping of correlated electrons $t_d$ appears not only inside $\chi_{cf}$,
but also on the left hand side of the equation, arising from the coupling between
nearest neighbour rotors in $H_\theta$ ($t_d e^{-i\theta_i}e^{i\theta_j}$).
This term competes with the interaction part ($\tsim U n_{\theta,i}^2$)
and tends to ``lock'' the angles of nearby rotors, therefore, in this second role, $t_d$ tends to 
enhance the appearance of a residue and therefore favors the destruction of the spin liquid
in favor of the appearance of the finite $\Phi$ leading to a metallic state.

\begin{figure}[b]
\centering
\includegraphics[width=0.45\textwidth]{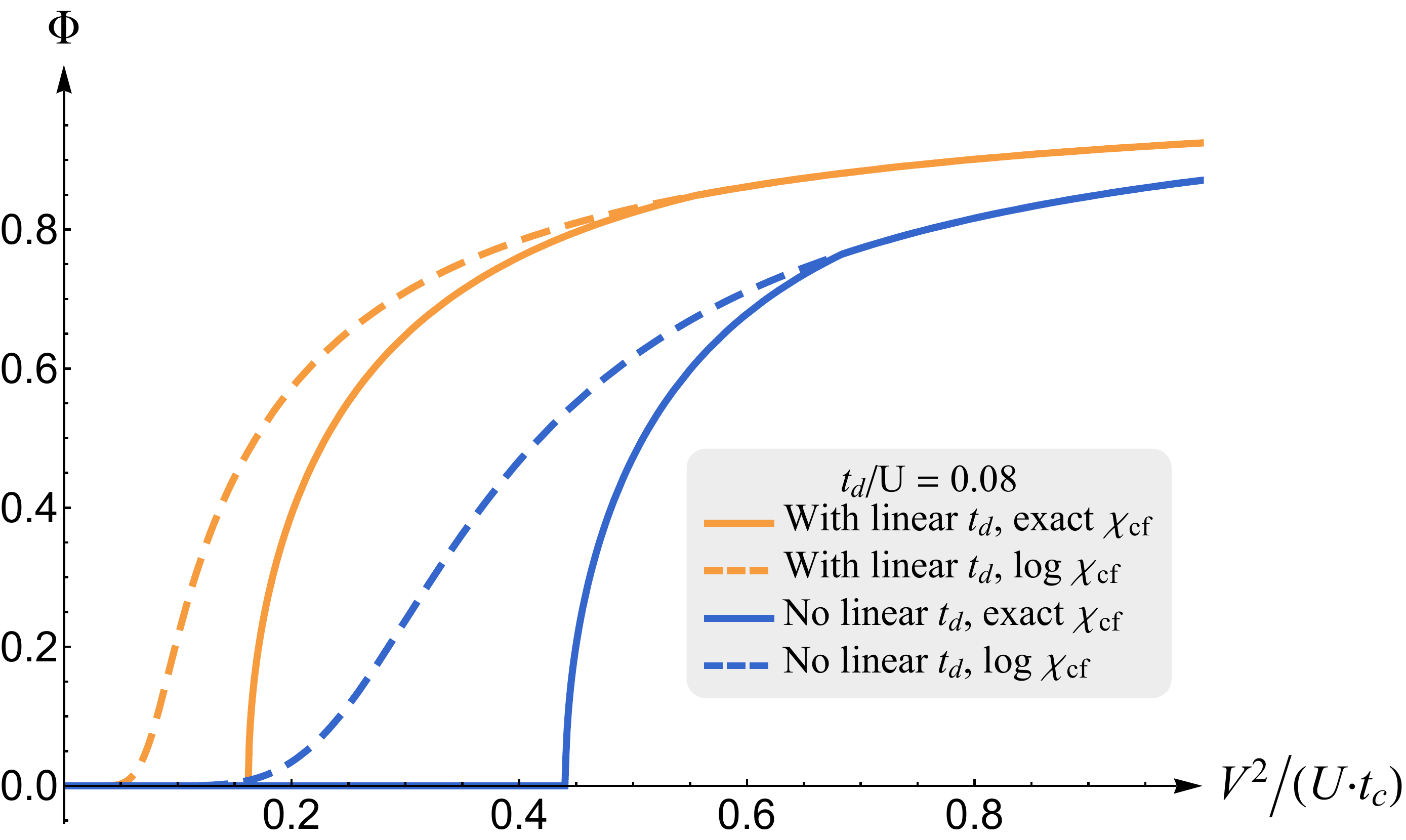}
\caption{
Solution of $\Phi$ for different types of self-consistent equations.
The orange lines stand for the self-consistent equation with the linear $t_d$ (nearest neighbour
coupling) term while the blue lines are for the case without the linear $t_d$ term.
The solid lines are for the case with exact form of $\chi_{cf}$ with a cut-off while the dashed
curves stand for the case with a (logarithmically) diverging $\chi_{cf}$ at small $V_f$.
The logarithmically diverging $\chi_{cf}$ always support a weak-coupling
instability to the heavy metal phase while for the exact $\chi_{cf}$, there
is a threshold of $V$ for the onset metallic phase.
The linear $t_d$ term in the left hand side of the self-consistent equation
will also help boost the heavy fermion phase, as expected.
}
\label{fig:phi-V-pp}
\end{figure}

To illustrate more concretely these contrasting roles of $t_d$ we compare the
solution of $\Phi$ as a function of $V^2/U$ for different types of \emph{modified}
self-consistent equations.
As shown by the dashed curves in \Fref{fig:phi-V-pp}, when the susceptibility $\chi_{cf}$ is
replaced by one which diverges logarithmically at small $V_f$ (dashed lines),
there is always a weak-coupling instability to the heavy fermion phase,
while for the exact $\chi_{cf}$ (solid lines), one has to reach a finite critical value of $V$ for the
occurrence of the heavy metal phase.
Moreover, when the linear $t_d$ terms from the left hand side of \Eqref{eq:self-cons}
is removed (blue lines), the heavy metal phase is also suppressed and one needs a larger $V$
to get a non-zero $\Phi$.

\begin{figure}
\centering
\includegraphics[width=0.45\textwidth]{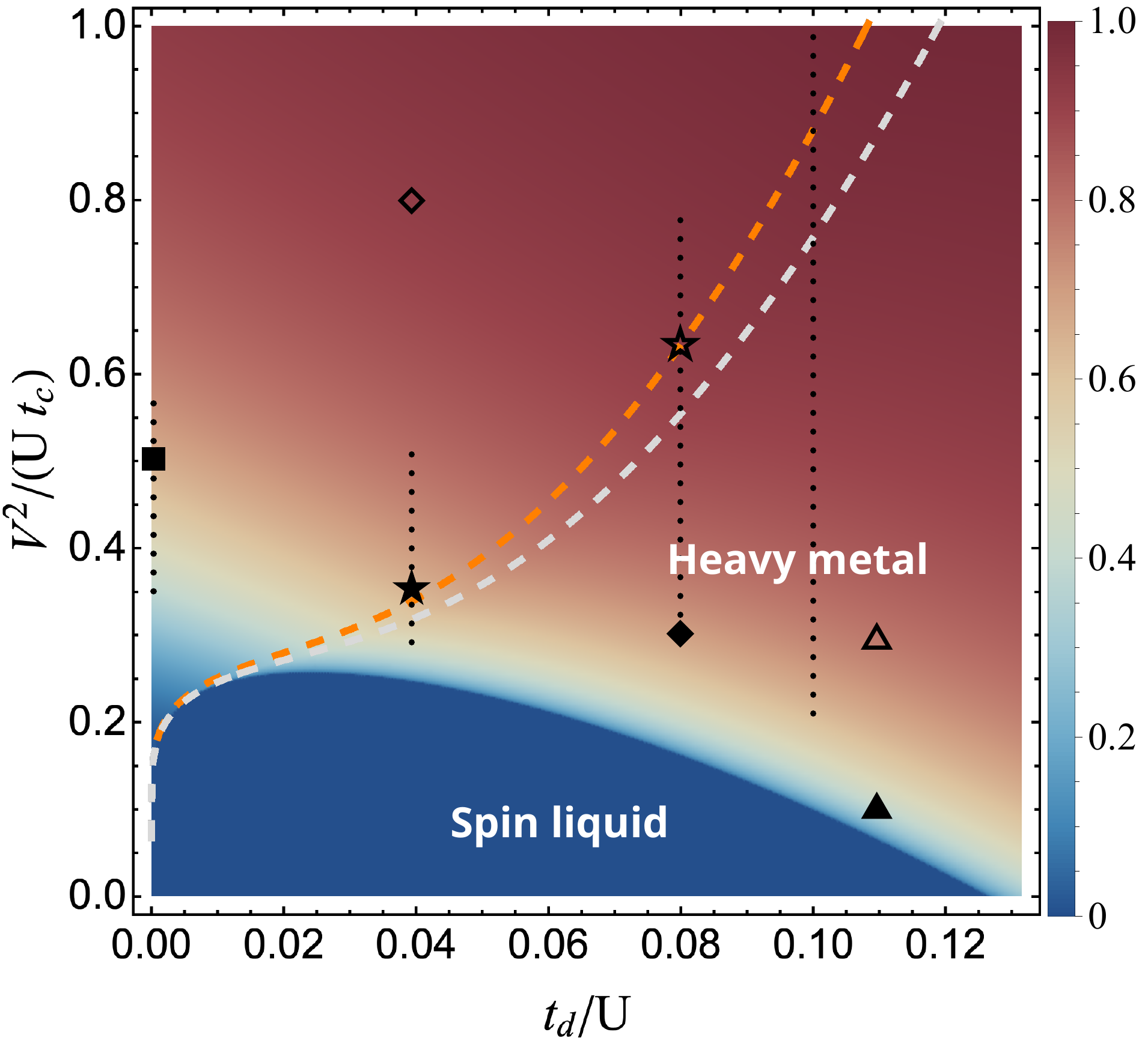}
\caption{
Phase diagram with $\xi \teq 1.2$ (density plot of $\Phi$).
The vertical scale is proportional to the Kondo coupling scale $J_K \tsim V^2/U$
while the horizontal scale is proportional to the hopping between the correlated electrons.
The dark blue region is the spin liquid with $\Phi \teq 0$ and
the light blue and red region stand for the heavy metal phase.
The gray dashed curve is the critical value of $V$ where the number of Fermi surfaces
of the system changes from two (below) to one (above) and the $\chi_{cf}$ changes from a 
constant plateau to a monotonically decreasing function of $V_f$ (see \Fref{fig:pb-chi-pp}(b)).
The orange dashed curve indicates where the two heavy quasiparticle bands 
develop an indirect band gap.
Dotted lines and symbols indicate where detailed LDOS spectra are calculated 
as a guiding reference for subsequent \Frefs{fig:ldos-anderson} -- \ref{fig:ldos-td-0.11}.
}
\label{fig:phase-diagram-pp}
\end{figure}

From the analysis above, one can see that either a very large $t_d$
(nearby rotors lock strongly) or a very small $t_d$
(susceptibility of the $c$-$f$ coupling diverges) will enhance the tendency towards
heavy Fermi liquid order and suppress the tendency towards the spin-liquid insulating phase.
This conclusion is further confirmed by the (zero temperature) phase diagram
\Fref{fig:phase-diagram-pp} obtained by explicitly solving the self-consistent equation
(the boundary in this phase diagram is the same previously shown in \Fref{fig:pb-chi-pp}(a)).
As can be seen from \Fref{fig:phase-diagram-pp},
the insulating spin liquid phase has a dome shape in the phase diagram,
which will be suppressed by very small or large $t_d$.
The gray dashed line indicates the critical value of $V$, above which $E_2$ band is fully occupied
and the metallic phase has a single Fermi surface.
The orange dashed line marks the boundary where the two heavy fermion bands start to develop
an indirect gap, which occurs for parameters above such orange line
(see further discussion in \Secref{sec:ldos}).
% ------------------------------------------------------------------------------
% FIGURE
% ------------------------------------------------------------------------------
% ------------------------------------------------------------------------------

% ------------------------------------------------------------------------------
% FIGURE
% ------------------------------------------------------------------------------
% ------------------------------------------------------------------------------
%
\subsection{Particle-hole dispersion}
In this section we discuss the results for the case where itinerant electrons are
hole-like which can be accounted for by simply changing $t_c \rightarrow -t_c$
in their energy dispersion
(\Eqref{eq:c-bands}).
\subsubsection{Phase-boundary}
\begin{figure}
\centering
\includegraphics[width=0.49\textwidth]{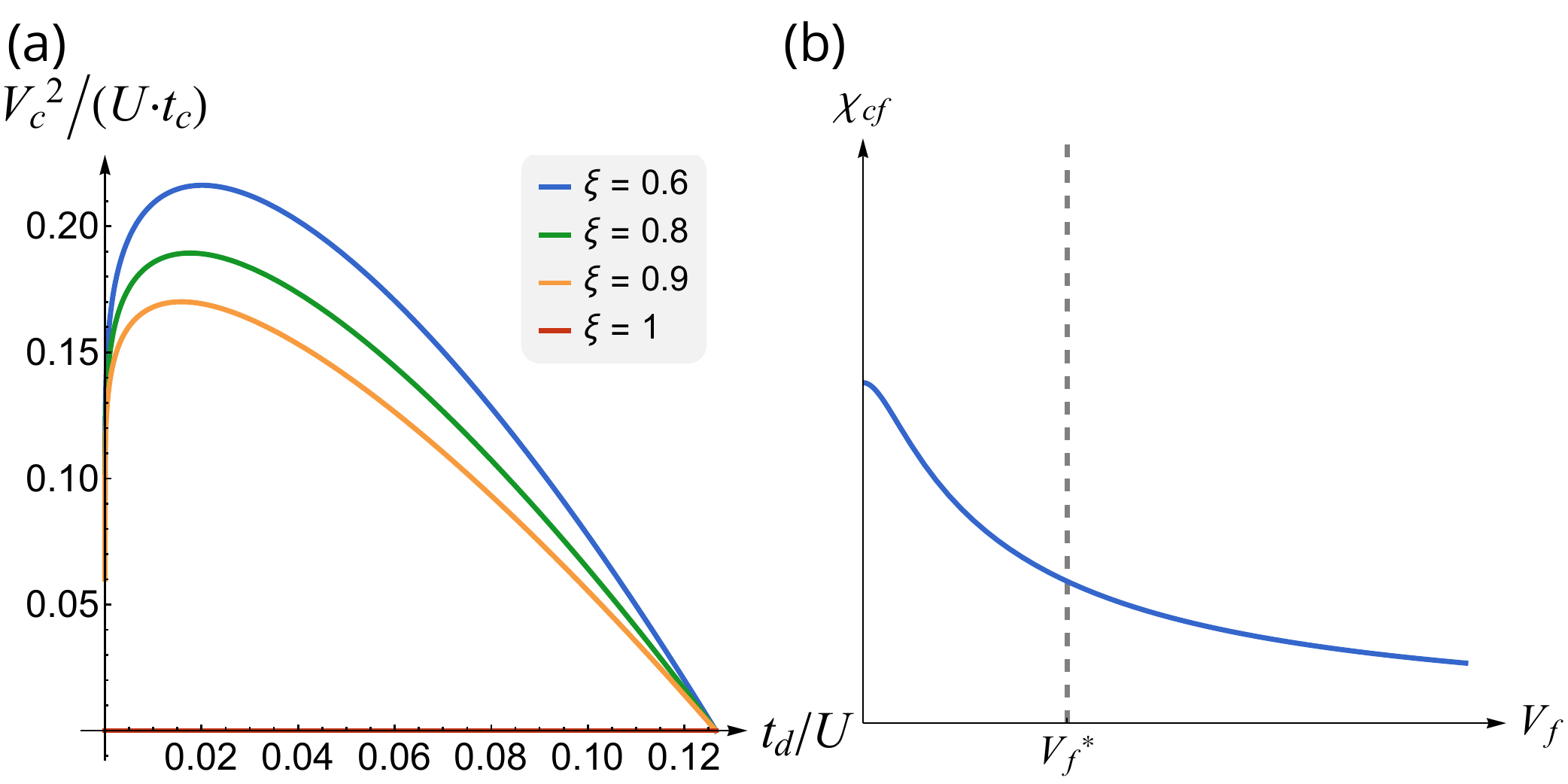}
\caption{
(a) Phase boundary for particle-hole dispersion at various filling of the metallic electrons.
As $\xi \rightarrow 1$, the spin liquid phase gets suppressed and at exactly half-filling of the
metal, it can exist only within the $V \teq 0$ line.
(b) $\chi_{cf}$ as a function of $V_f$ for the particle-hole dispersion with
$\xi \teq 0.6,\ T_f \teq 0.1 t_c$. Similar to the particle-particle case, $\chi_{cf}$ is a decreasing
function of $V_f$.
}
\label{fig:pb-chi-ph}
\end{figure}

When the metallic electron's band structure is hole-like,
the susceptibility $\chi_{cf}$ will have a stronger $\xi$ dependence compared to the
particle-particle case.
It can be shown that within the spin liquid phase ($V_f \teq 0$), it is given by:
\begin{align}
\chi_{cf}^{(0)} &= \frac{2}{3\Lambda^2 (T_f+t_c)}
\ln \left( \frac{( T_f/t_c+\xi )( T_f/t_c+2-\xi )}{T_f/t_c(1-\xi)^2} \right).
\end{align}
Thus for $\xi \teq 1$, i.e., when both the itinerant electrons and spinons are at
half-filing, the two bands are perfectly \emph{nested}, the band structure leads to a divergent
susceptibility $\chi_{cf}$ for all values of $t_d$,
which indicates that the spin liquid is unstable against a transition into the
Kondo insulating phase at arbitrarily small $V$.
Figure~\ref{fig:pb-chi-ph}(a) shows the phase boundary between the spin liquid
and the heavy fermion metallic phase. Similar to the particle-particle case, as $t_d \rightarrow 0$,
the critical value of
$J_K \tsim V^2/U$ decreases logarithmically with $t_d$. Moreover, for the particle-hole
case, the phase boundary now also has a $\xi$-dependence, as expected from the
$\xi$-dependence of $\chi_{cf}^{(0)}$.
As $\xi \rightarrow 1$, the spin liquid phase is suppressed and when $\xi \teq 1$,
it only exists along the $V \teq 0$ line \Fref{fig:pb-chi-ph}(a).
It should be noted that at $V \teq 0$, the critical $t_d/U$ for the Mott transition is always the
same ``universal'' value around $1/8$, this is because the $d$ and $c$ electrons are decoupled
in this case and the problem reduces to the metal to insulator transition for the triangular
lattice Hubbard model.
% ------------------------------------------------------------------------------
% FIGURE
% ------------------------------------------------------------------------------
% ------------------------------------------------------------------------------
\subsubsection{Turning on of the heavy fermion phase}
\begin{figure}
\centering
\includegraphics[width=0.45\textwidth]{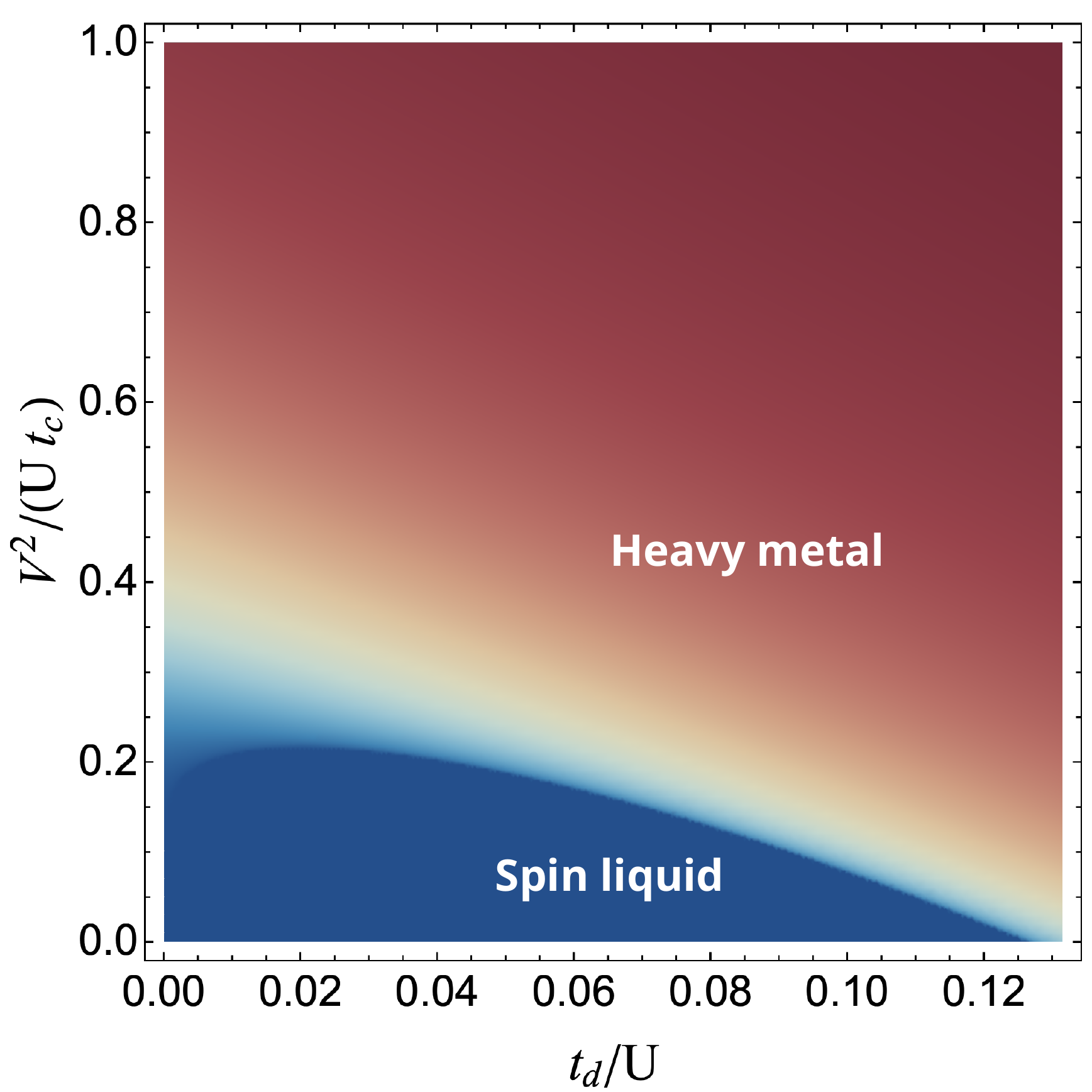}
\caption{
Phase diagram for the particle-hole case with $\xi \teq 0.6$.
The spin liquid phase has a dome shape and the phase boundary has qualitatively the
same behaviour as the particle-particle case.
}
\label{fig:pd-ph-xi-0.6}
\end{figure}

For the case with $\xi \tlt 1$, weakly inside the heavy-fermion metallic phase, where
the quasi-particles' energy dispersion $E_{1,k}$ and $E_{2,k}$ has the 
Mexican-hat shape,
it turns out that in order to maintain the half-filling constraint of the spinon,
we find that $E_{2,k}$ band is fully filled
while the $E_{1,k}$ band is partially occupied and features two Fermi surfaces, as shown by the green dashed lines in \Fref{fig:bands}(b).
The $\mu_F$ can be solved from $\langle f_{i,\sigma}^{\dagger} f_{i,\sigma} \rangle \teq 1/2$ and
the $\chi_{cf}$ as a function of $V_f$ can be obtained accordingly.
Similar to the particle-particle case, at finite $t_d$, $\chi_{cf}$ tends to saturate as
$V_f \rightarrow 0$ and it is diverging in the atomic limit ($t_d \rightarrow 0$).
For rather large $V_f$, $E_{1,\Lambda}$ becomes smaller than $0$ and there is only one
Fermi surface for the system (see the orange dashed lines in \Fref{fig:bands}(b)).
A plot of $\chi_{cf}$ at $\xi \teq 0.6$ is shown in \Fref{fig:pb-chi-ph}(b), as expected,
it is a decreasing function of $V_f$.
The phase diagram for this case is shown in \Fref{fig:pd-ph-xi-0.6}.

\begin{figure}
\centering
\includegraphics[width=0.45\textwidth]{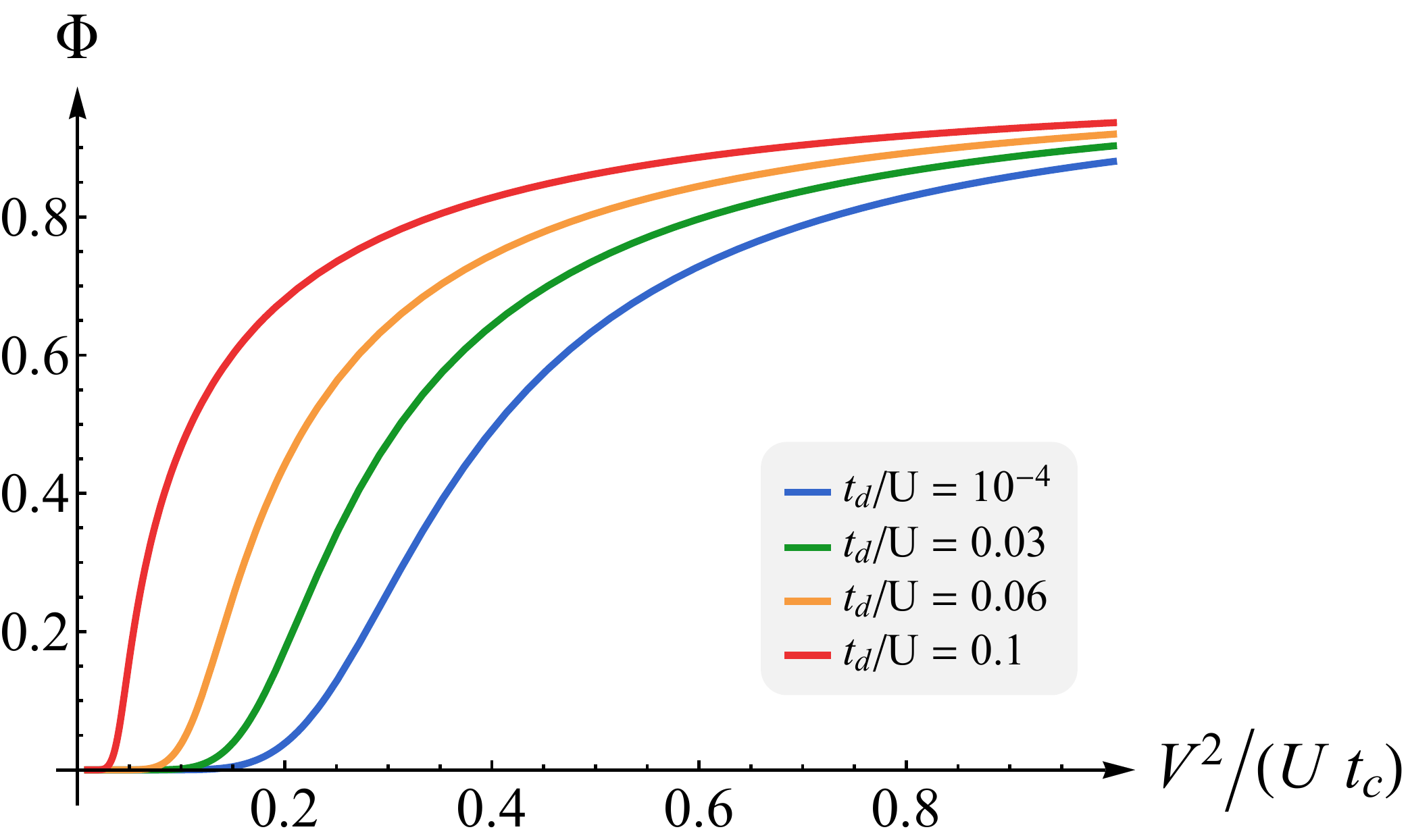}
\caption{
$\Phi$ as a function of $V^2/U$ for $\xi \teq 1$ at different value of $t_d/U$.
As expected, the metallic phase turns on in the form of a weak coupling instability with
$V$.
}
\label{fig:phi-ph-xi-1}
\end{figure}

\begin{figure}
\centering
\includegraphics[width=0.45\textwidth]{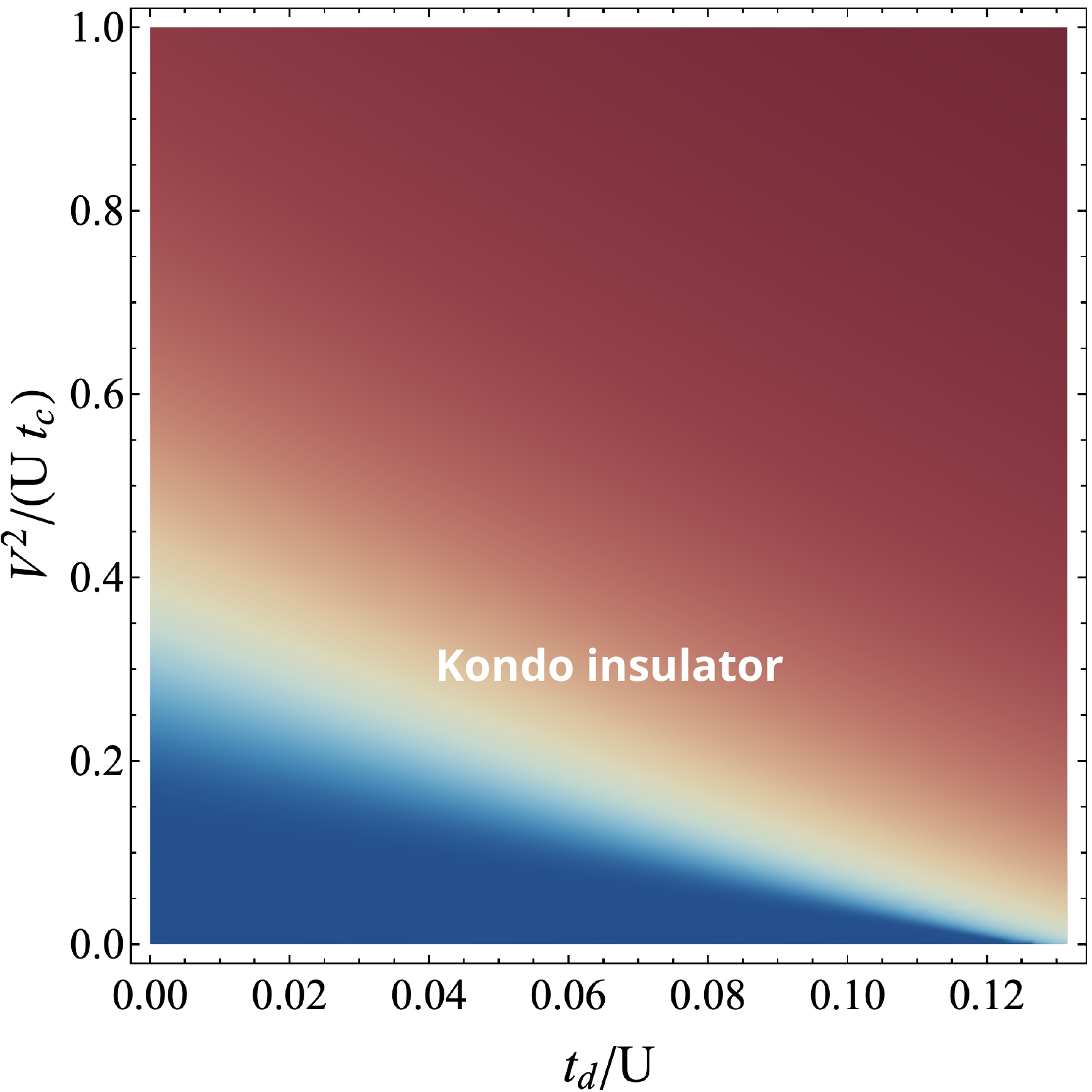}
\caption{
Phase diagram for the particle-hole case with perfect nesting ($\xi \teq 1$).
The system is in Kondo insulating at any finite $V$ since the Fermi surface of the heavy
electrons are fully gapped out, and the spin liquid phase exists strictly only at the $V \teq 0$ line.
}
\label{fig:pd-ph-xi-1}
\end{figure}

As for the special case when $\xi \teq 1$, as explained before, because the spinon and the
itinerant electron bands are nested in this case, the susceptibility $\chi_{cf}$ diverges as
$V_f \rightarrow 0$.
As a result, one expects a weak coupling instability
from the spin liquid state to that with heavy electrons.
Notice however that this state is not a metal but a Kondo insulator,
since the Fermi surfaces are completely gapped out by the hibridization due to the perfect
nesting.
As can be seen from \Fref{fig:phi-ph-xi-1}, the Kondo insulating phase turns on more
rapidly for larger $t_d/U$.
The phase diagram for this case is shown in \Fref{fig:pd-ph-xi-1}.
% ------------------------------------------------------------------------------
% FIGURE
% ------------------------------------------------------------------------------
% ------------------------------------------------------------------------------
% ------------------------------------------------------------------------------
% FIGURE
% ------------------------------------------------------------------------------
% ------------------------------------------------------------------------------
% ------------------------------------------------------------------------------
% FIGURE
% ------------------------------------------------------------------------------
% ------------------------------------------------------------------------------
\section{Tunnelling DOS} \label{sec:ldos}
In the recent experiment by Ruan et al. \cite{Wei2020}, a monolayer \TaSe ,
which is originally an insulator, is placed on top of a metallic monolayer 1H-TaSe\textsubscript{2}.
The system was studied by STM, where the tip is primarily coupled to the top layer (\TaSe).
Surprisingly, a narrow peak around zero bias was found. It was found that this coherent
peak can be broadened by increasing temperature and the temperature dependence
of its width can be fitted to a form (see \Eqref{eq:width-exp1}) which describes the Kondo 
resonance for the single impurity Kondo problem (as shown in the Fig. 2(c) of Ref.~\cite{Wei2020}).
This observation was then taken as an indication of the existence of the local magnetic moment
in the \TaSe\ layer, which couples to the metallic substrate (the 1H layer).
Combining this with the further observation of a real space modulation of the 
electronic structure, it was suggested that the pristine \TaSe\ monolayer is 
likely to host the QSL phase.

This motivates us to study if this behaviour could also appear in our theoretical model,
e.g., in certain regimes of the heavy metal phase.
In this section, we discuss the behaviour of the LDOS of the correlated
$d$ electrons in the metallic phase, which is the quantity reflected by the STM $dI/dV$ curve.
The thermal Green function of the $d$ electron can be written as:
\begin{align}
G_d(\tau,\mathbf{r}) & = -\langle T_\tau d_{\mathbf{R} + \mathbf{r}}(\tau)
d_{\mathbf{R}}^{\dagger}(0)
\rangle \\ \nonumber
& = G_f(\tau, \mathbf{r}) G_{\theta} (\tau,\mathbf{r}),
\end{align}
here $G_f(\tau, \mathbf{r})$ and $G_{\theta} (\tau,\mathbf{r})$ are Green functions of the
spinon and rotor, with the definition:
\begin{subequations}
\begin{align}
G_f(\tau, \mathbf{r}) & = -\langle T_\tau f_{\mathbf{R} + \mathbf{r}}(\tau)
f_{\mathbf{R}}^\dagger (0) \rangle, \\
G_{\theta} (\tau,\mathbf{r}) & = \langle T_\tau e^{i \theta_{\mathbf{R} + \mathbf{r}}(\tau)}
e^{-i \theta_{\mathbf{R}}(0)} \rangle.
\end{align}
\end{subequations}
As pointed out from previous studies \cite{Florens2004,Zhao2007}, the Matsubara Green function
of $d$ electrons can be separated into a \emph{coherent} part and an \emph{incoherent} part:
\begin{subequations}
\begin{align}
G_d(i\omega_n,\mathbf{r})  &= G^{coh}_d(i\omega_n,\mathbf{r})
+ G^{inc}_d(i\omega_n,\mathbf{r}), \\
G^{coh}_d(i\omega_n,\mathbf{r}) &= \Phi^2 G_f(i\omega_n,\mathbf{r}).
\end{align}
\end{subequations}
The coherent part is mainly peaked at $\omega \tsim 0$ while the incoherent part
captures features at larger energy scales $\omega \tsim U$.
In this work, we are mainly interested in the feature of LDOS near $\omega \teq 0$ and we will
focus on the coherent part.
From the slave rotor mean-field theory, since the fermionic part of the Hamiltonian is 
non-interacting, it can be shown that the Matsubara Green function of spinon has the form:
\begin{equation}
G_f(i\omega_n, k) = \cos^2(\alpha_k) G_1(i\omega_n, k) + \sin^2(\alpha_k) G_2(i\omega_n, k),
\end{equation}
where $G_{1/2}(i\omega_n,k) = 1/\left( i\omega_n - E_{1/2,k} \right)$ are the Green
function of the self-consistent band-diagonal quasi-particles that result from the coherent
mixing of the correlated and the itinerant electron.
By analytical continuation, the spectral function of the spinons can be obtained:
\begin{align}
A_f(\omega,k) & = -\frac{1}{\pi} \Im G_f( \omega+i 0_+, k ) \\ \nonumber
& = \cos^2(\alpha_k) \delta(\omega - E_{1,k})
+ \sin^2(\alpha_k) \delta(\omega - E_{2,k}), \label{eq:spectal-function}
\end{align}
and the LDOS for the spinon
$\rho_f(\omega) \teq \frac{1}{N} \sum_k A_f(\omega,k)$ can be obtained accordingly.
\subsection{Zero temperature mean-field LDOS}
We are particularly interested in understanding the tunnelling density of states for
experiments in \TaSe\ where the dispersion of itinerant electron is likely
to be particle like. Here we explored in detail the particle-particle case and
we take the bare band filling of the itinerant electrons to be
$\xi = 1.2$ (this value is taken arbitrarily as the physics should not be very sensitive
to the detailed value of $\xi$).
We are mainly focused on three regimes:
i) Anderson limit with $t_d \teq 0$, ii) moderate $t_d$ along the orange dashed line in
\Fref{fig:phase-diagram-pp}, iii) large $t_d$ near the metal-insulator transition of Hubbard model.

\begin{figure}
\centering
\includegraphics[width=0.49\textwidth]{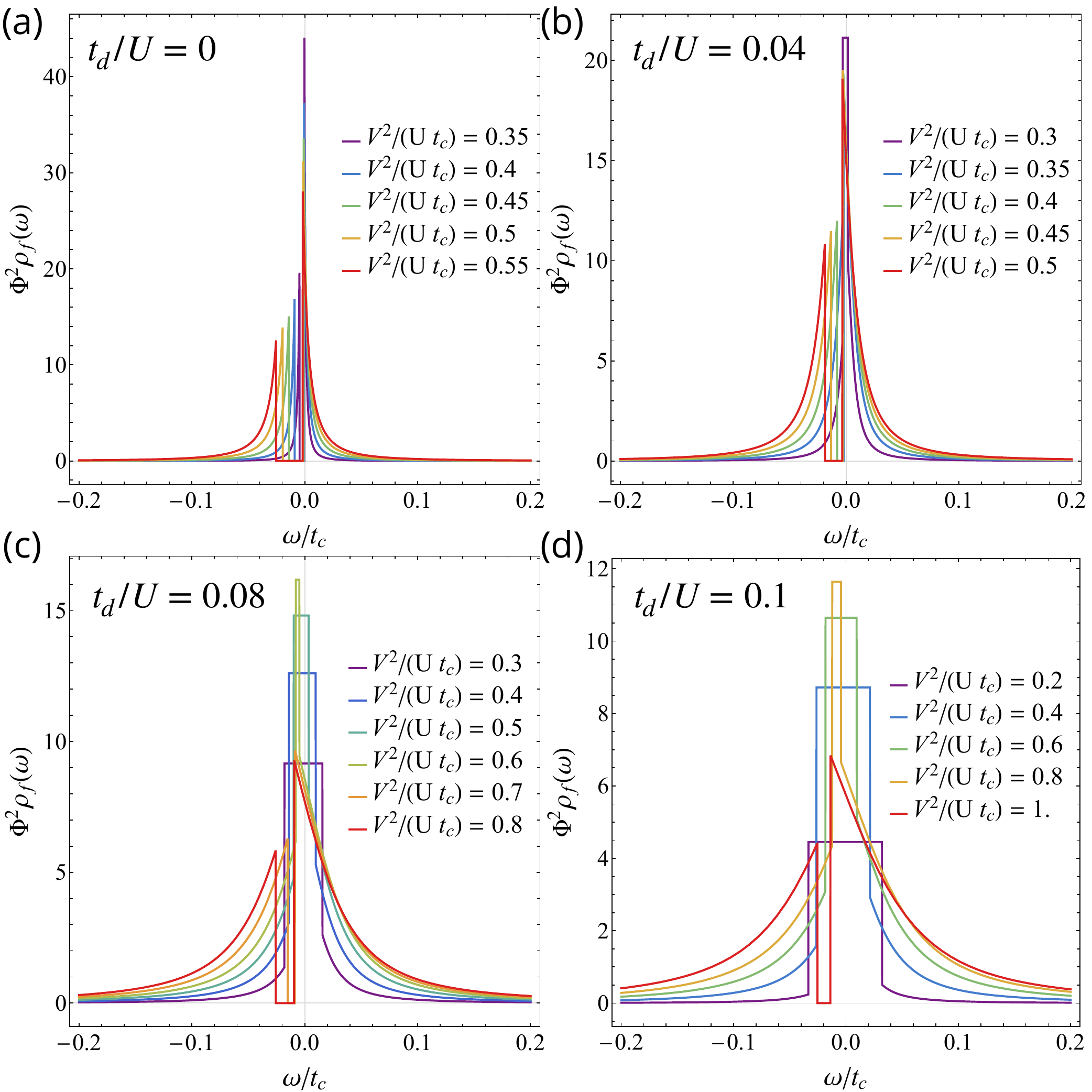}
\caption{
Mean-field LDOS without disorder and quasiparticle life-time broadening effects for the case of
(a) $t_d/U \teq 0$, (b) $t_d/U \teq 0.04$, (c) $t_d/U \teq 0.08$ and (d) $t_d/U \teq 0.1$.
Within each case, the Kondo coupling $J_K \tsim V^2/U$ is increased gradually
(along the black dotted lines in \Fref{fig:phase-diagram-pp}).
In the Anderson limit, it is clear that within the heavy metal phase, there is a coherent gap
opened below the Fermi level.
On the other hand, when $t_d/U$ is finite, the spinon band is dispersive with a finite bandwidth.
So for small $J_K$, the band dispersion of heavy quasiparticles are still overlapping with each
other (see the green dashed lines in \Fref{fig:bands}(a)), and leads to a plateau like LDOS at
small $\omega$. When $J_K$ is large and above the orange dashed line in the phase 
diagram (see \Fref{fig:phase-diagram-pp}), the two heavy quasiparticle bands are fully separated
in energy and the LDOS exhibits a gap between the two peaks.
}
\label{fig:ldos-bare}
\end{figure}

\Fref{fig:ldos-bare} shows the zero temperature mean-field LDOS of correlated $d$ electrons
at different regimes of the phase diagram, as indicated by the black dotted lines in
\Fref{fig:phase-diagram-pp}.
In the Anderson limit (see \Fref{fig:ldos-bare}(a)), the mean-field LDOS opens a coherent
band gap enhanced by increasing the Kondo coupling $J_K$,
which is the expected behaviour for the periodic Anderson model.
When $t_d/U$ is finite (see \Frefs{fig:ldos-bare}(b), (c) and (d)),
the spinon acquires a band dispersion.
Consequently, when $\Phi$ is small at small $J_K$, the quasiparticle bands are still 
overlapping with each other in energy (see green dashed line in \Fref{fig:bands}(a))
and the LDOS shows a plateau-like peak near $\omega \tsim 0$. The width of the plateau is given
mainly by the spinon bandwidth.
As $J_K$ becomes larger, the overlap between the two bands decreases and the width of the flat
peak is reduced. At some intermediate scale marked by the
orange dashed line in \Fref{fig:phase-diagram-pp}, the Kondo coupling and the Heisenberg
exchange interaction compete, resulting in a narrow peak whose width is much less than $J_K$
or $J_H$ inidividually. Finally, when $J_K$ is greater than a critical value indicated by the orange
dashed line in
\Fref{fig:phase-diagram-pp}, the two quasiparticle bands become fully separated and
the LDOS behaves similarly to the Anderson limit with a finite gap sandwiched by two
peaks.
As can be seen clearly, near the the metal-insulator transition of the Hubbard model, the LDOS peak is much broader than in the small $t_d/U$ limit.
It should be noted that the perfect flatness of the peak is
an artefact of parabolic band dispersion adopted in our study,
and a more realistic tight-binding model would give rise to a dispersive peak.
Below we will describe how these LDOS features
are broadened by temperature and by extrinsic disorder effects.
\subsection{Broadening due to finite temperature and disorder}
At finite temperature the tunneling conductance is given by the LDOS convolved with the thermal
broadening due to the thermal distribution of electrons in the lead. This effect has been removed in 
the  experiment
\cite{Nagaoka2002} and we also do not include it in our theory.
After removing this, it is notable that  the experiment
shows a single peak which can be fitted with a Lorentzian with a temperature dependent half
maximum half width:
\begin{equation} \label{eq:width-exp1}
\Gamma_{exp} = \sqrt{2T_K^2 + \pi^2 T^2},
\end{equation}
This form of the width was found in an earlier experiment that detected the Kondo peak in a single
impurity and has been considered a signature of the single impurity Kondo
problem \cite{Nagaoka2002}.
The low temperature width therefore allows to extract $T_K$ from experiments.
Furthermore, at large temperatures compared to $T_K$ the width scales approximately as
$\pi\,T$, which places a constraint on the theory. We have re-examined the theoretical basis of
\Eqref{eq:width-exp1} and came to the conclusion that while the
derivation given in \cite{Nagaoka2002} is not well justified
and there is a small correction to the width from \Eqref{eq:width-exp1} at low temperatures,  it does provide
a correct value of the slope of $\Gamma$-$T$ curve at high temperatures, which is $\pi$.
Details are given in the \Appref{sec:append-ldos}.
In this work we do not fit the experimental data to the 
single impurity Kondo problem, but rather to the periodic Anderson-Hubbard model. As we shall see below, by introducing a Fermi liquid type quasiparticle life-time together 
with a disorder induced width, it is possible to fit the data in certain parameter ranges.

As it is well known from the theory of single Kondo impurity and Kondo lattice problems
\cite{Read1983,Read1983a,Coleman1984,Auerbach1986},
the fluctuations around the mean-field configuration which give rise to quasi-particle interactions, 
lead to a characteristic temperature and frequency dependent quasi-particle lifetime.
In order to account for these effects, we add the following semi-phenomenological imaginary
part to the quasi-particle self-energy \cite{Zheng1996}:
\begin{equation} \label{eq:sigma-fl}
\Sigma_{FL}( \omega, T ) = -i \frac{1}{2\pi E_0} ( \omega^2 +  (\pi k_B T)^2 ).
\end{equation}
In addition to this intrinsic quasi-particle interaction lifetime, disorder is another important
agent in broadening the density of states in experiments, and we account for this by adding
a constant impurity scattering rate $\gamma_0$ into the imaginary part of the self-energy,
as follows:
\begin{subequations}
\begin{align}
G_{1/2}( \omega + i 0_+, k ) & = \frac{1}{\omega - E_{1/2,k} - \Sigma( \omega, T ) } \\
\Sigma( \omega, T ) & = -i \gamma_0 + \Sigma_{FL}( \omega, T ). \label{eq:self-energy-full}
\end{align}
\end{subequations}
It  should be noted that the energy scale $E_0$ controlling the quasi-particle interaction effects
in \Eqref{eq:sigma-fl}, is usually of the 
order of the bandwidth for a normal Fermi liquid (large $t_d$),
while for a Kondo lattice ($t_d \teq 0$), it is of the order of the Kondo temperature
$T_K \tsim 2 V_f^2 / D_c$ with $D_c$ being the half bandwidth of itinerant electrons.
In order to capture both regimes, we use a phenomenological expression of $E_0$
that interpolates between these two limits, as follows:
\begin{equation} \label{eq:E_0}
E_0 = \sqrt{ T_K^2 + W_{sp}^2 },
\end{equation}
with $W_{sp}$ being the spinon bandwidth.
% ------------------------------------------------------------------------------
% FIGURE
% ------------------------------------------------------------------------------
% ------------------------------------------------------------------------------
% ------------------------------------------------------------------------------
% FIGURE
% ------------------------------------------------------------------------------
\begin{figure}
\centering
\includegraphics[width=0.49\textwidth]{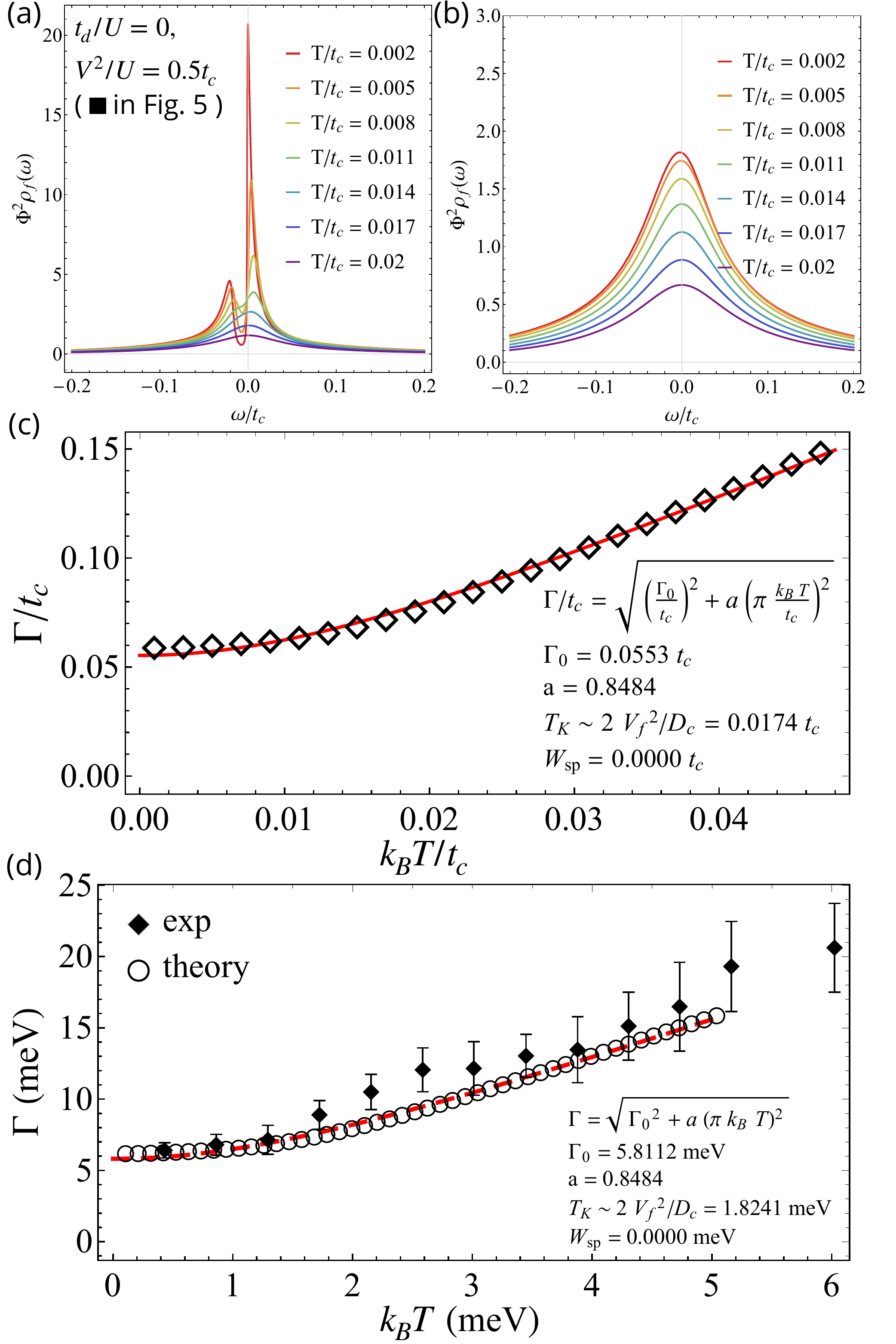}
\caption{
LDOS for the particle-particle case ($\xi \teq 1.2$) with $t_d/U \teq 0, V^2/U \teq 0.5\, t_c$
(indicated by $\filledmedsquare$ in \Fref{fig:phase-diagram-pp}).
(a) LDOS with the self-energy being $\Sigma_{FL}( \omega, T )$ only.
It is clear that in the low temperature limit, the spectral function has the two-peak behaviour
at $\omega \tsim 0$, which is due to the opening of a band gap in the dispersion of heavy quasiparticles. This is the signature of a coherent heavy Fermion band 
in the kondo lattice problem.
At higher temperature, there is only a single peak around $\omega \tsim 0$
due to the broadening effects in $\Sigma_{FL}( \omega, T)$.
(b) LDOS for self-energy from \Eqref{eq:self-energy-full} with $\gamma_0 \teq 0.05\, t_c$.
In this case the disorder effect ($\gamma_0$ term) is able to broaden the LDOS and changes it
into a single-peak.
(c) Width in unit of $t_c$.
(d) Fitting to experimental data (extracted from Ref.~\cite{Wei2020}) with $t_c \teq 105 \meV$.
The experimental data can be well fitted by the theoretical result.
}
\label{fig:ldos-anderson}
\end{figure}
% ------------------------------------------------------------------------------

As mentioned above, in the Anderson limit, the mean-field LDOS will
have two peaks separated by the gap.
However, once the self-energy is included, the mean-field spectral
function will be broadened and it is possible to obtain a single-peak behaviour.
This can be seen clearly from \Fref{fig:ldos-anderson}, which shows the case of
$t_d/U \teq 0,\ V^2/U \teq 0.5 t_c$  (as indicated by the $\filledmedsquare$ in
\Fref{fig:phase-diagram-pp}).
By including only the $\Sigma_{FL}$
(see \Fref{fig:ldos-anderson}(a)), at very low temperatures, the LDOS has two peaks
separated by a band gap. When a finite impurity scattering rate (here we take
$\gamma_0 \teq 0.05\, t_c$) is taken into account, the LDOS is broadened into a single-peak, as shown in \Fref{fig:ldos-anderson}(b).
We further calculated the half maximum half width of LDOS at different temperatures and
compare it with the experimental results.
We fit our theoretical data with a function of the form
\begin{equation} \label{eq:width-fit}
\Gamma = \sqrt{ \left( \Gamma_0 \right)^2 + a \pi^2 \left( k_B T \right)^2 },
\end{equation}
which is expected for the single-impurity Anderson model \cite{Yamada1975,Costi1994}.
Previous theoretical works find that the experimental data can be well fitted with $a \tapprox 1$.
According to our theoretical calculation, for the case with $V^2/U \teq 0.5\, t_c$ and
$\gamma_0 \teq 0.05\, t_c$, the data can be well fitted with $a \tapprox 0.85$, as can be seen
from \Fref{fig:ldos-anderson}(c), where all quantities are presented in unit of $t_c$.
Nevertheless, once we take $t_c \teq 105 \meV$ so that the lowest temperature width matches
with the experimental one, we also find quantitatively good fit to the experimental result.
In other words, the experimental data can be described by a periodic Anderson model with a
finite impurity scattering rate.

\begin{figure*}
\centering
\includegraphics[width=0.98\textwidth]{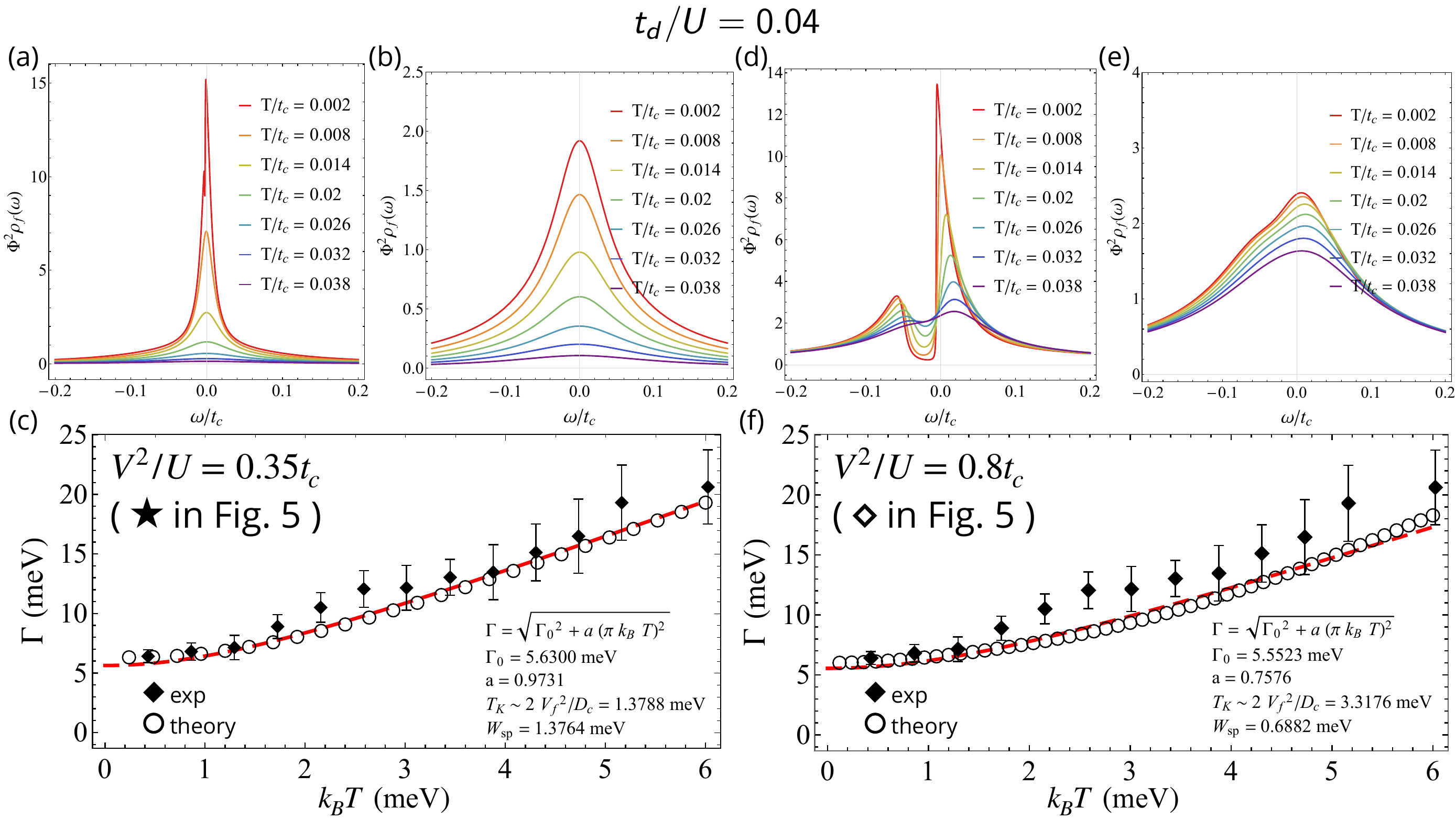}
\caption{
LDOS at $t_d/U \teq 0.04$.
(a)-(c) With $V^2/U \teq 0.35\, t_c$ (indicated by $\filledstar$ in
\Fref{fig:phase-diagram-pp}).
(a)-(b) LDOS without/with $\gamma_0$ in the self-energy.
(c) Width fitted to the experiment with $t_c \teq 120 \meV$.
The experimental data can be relatively well fitted by this case.
(d)-(f) With $V^2/U \teq 0.8\, t_c$ (indicated by $\diamond$ in
\Fref{fig:phase-diagram-pp}).
(d)-(e) LDOS without/with impurity scattering in the self-energy.
(f) Fitting of the width to experiment with $t_c \teq 60 \meV$.
This case is much above the orange dashed line in \Fref{fig:phase-diagram-pp} and
the two quasiparticle bands are separated form each other.
}
\label{fig:ldos-td-0.04}
\end{figure*}

\begin{figure*}
\centering
\includegraphics[width=0.98\textwidth]{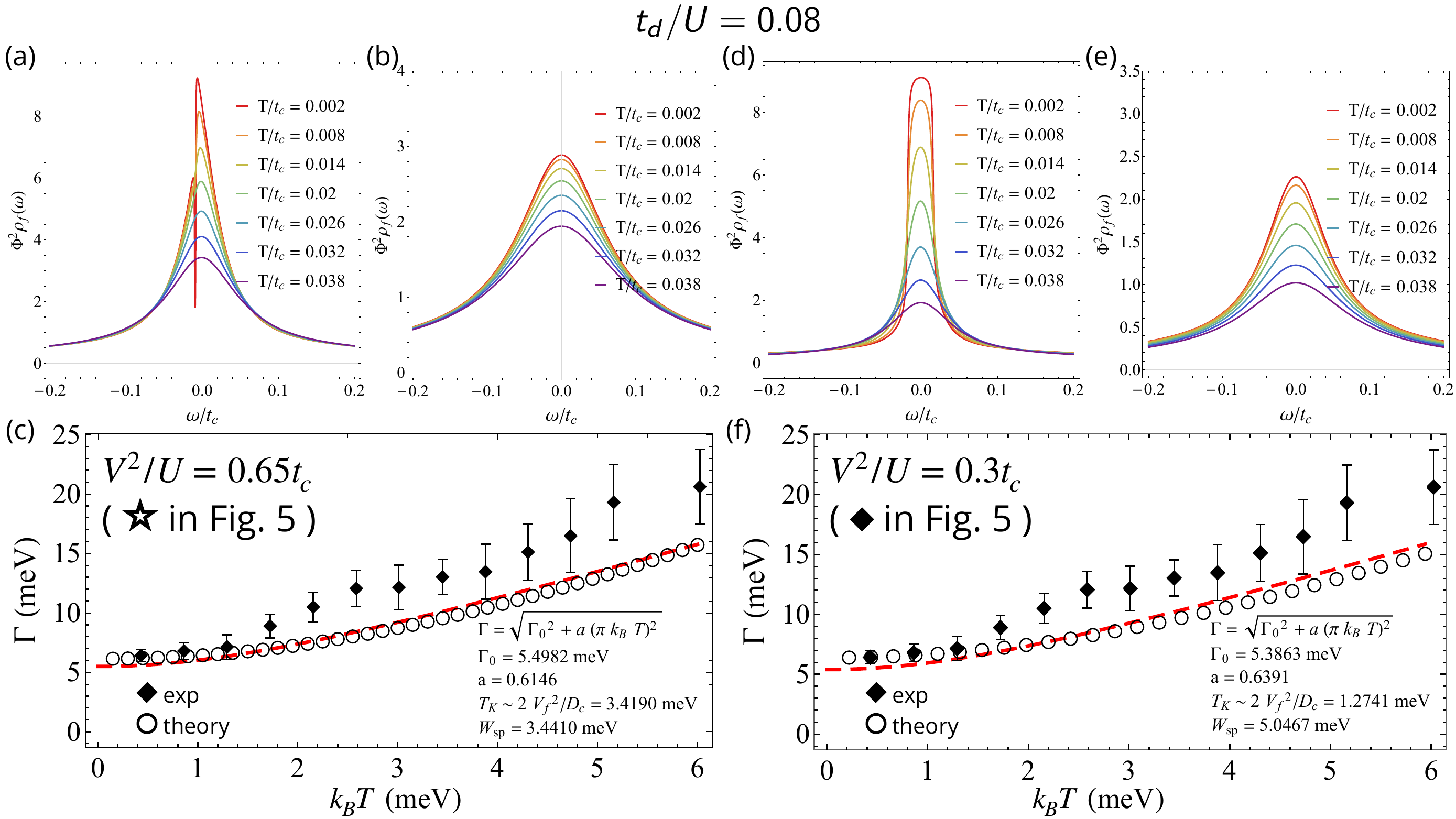}
\caption{
LDOS at $t_d/U \teq 0.08$.
(a)-(c) With $V^2/U \teq 0.65\, t_c$ (indicated by $\medstar$ in
\Fref{fig:phase-diagram-pp}).
(a)-(b) LDOS without/with $\gamma_0$ in the self-energy.
(c) Fitting of the width to experiment with $t_c \teq 75 \meV$.
In this case the theory lies below the data because the slope $a$ is becoming too small.
(d)-(f) With $V^2/U \teq 0.3\, t_c$ (indicated by $\filleddiamond$ in
\Fref{fig:phase-diagram-pp}).
(d)-(e) LDOS without/with impurity scattering in the self-energy.
(f) Fitting of the width to experiment with with $t_c \teq 110 \meV$.
This case is below the orange dashed line and the two quasiparticle bands overlaps.
}
\label{fig:ldos-td-0.08}
\end{figure*}

When $t_d$ is finite, as shown in the mean-field results above,
one expects to see either a plateau-like peak (with small $J_K$) or a finite
gap sandwiched by two peaks (rather large $J_K$) in the LDOS.
In any case, the inclusion of a finite imaginary self-energy can broaden the curve.
Along the orange line, since the two mean-field bands of heavy quasiparticles are about to
separate, the LDOS of spinon should have only a single peak around $\omega \tsim 0$.
\Frefs{fig:ldos-td-0.04}(a)-(c) and \ref{fig:ldos-td-0.08}(a)-(c) show two points
close to the line:
$t_d/U \teq 0.04$, $V^2/U \teq 0.35\, t_c$ and $t_d/U \teq 0.08$, $V^2/U \teq 0.65\, t_c$
(indicated by $\filledstar$ and $\medstar$ respectively in \Fref{fig:phase-diagram-pp}), it is
clear that the LDOS has only a single peak at $\omega \tsim 0$.
We find that the width as a function of temperature can also be relatively well
fitted by \Eqref{eq:width-fit}. To compare with the experimental data, as we did for the
Anderson limit, one can tune $t_c$ so that at the lowest temperatures the width is
consistent with the experimental one.
\Fref{fig:ldos-td-0.04}(c) and \Fref{fig:ldos-td-0.08}(c) show
the comparison of the width between the theoretical and
experimental results. $t_c$ is taken to be $120 \meV$ and $75 \meV$ separately.
We can see that the small spinon hopping case $t_d/U \teq 0.04$ can give rise to a
good fit to the experimental data. For the larger $t_d$ case ($t_d/U \teq 0.08$)
the  fit deteriorates because the coefficient $a$ is becoming too small.
% ------------------------------------------------------------------------------
% FIGURE
% ------------------------------------------------------------------------------
% ------------------------------------------------------------------------------
% ------------------------------------------------------------------------------
% FIGURE
% ------------------------------------------------------------------------------
% ------------------------------------------------------------------------------

%% ------------------------------------------------------------------------------
%% FIGURE
%% ------------------------------------------------------------------------------
%\begin{figure}
%\centering
%\includegraphics[width=0.5\textwidth]{Fig-13}
%\caption{
%LDOS at $t_d/U \teq 0.04,\ V^2/U \teq 0.8\, t_c$ (indicated by $\diamond$ in
%\Fref{fig:phase-diagram-pp}).
%(a)-(b) LDOS without/with impurity scattering in the self-energy.
%(c) Fitting of the width to experiment with $t_c \teq 60 \meV$.
%This case is much above the orange dashed line in \Fref{fig:phase-diagram-pp} and
%the two quasiparticle bands are separated form each other.
%}
%\label{fig:ldos-off-1}
%\end{figure}
%% ------------------------------------------------------------------------------

%% ------------------------------------------------------------------------------
%% FIGURE
%% ------------------------------------------------------------------------------
%\begin{figure}
%\centering
%\includegraphics[width=0.45\textwidth]{Fig-14}
%\caption{
%LDOS at $t_d/U \teq 0.08,\ V^2/U \teq 0.3\, t_c$ (indicated by $\filleddiamond$ in
%\Fref{fig:phase-diagram-pp}).
%(a)-(b) LDOS without/with impurity scattering in the self-energy.
%(c) Fitting of the width to experiment with with $t_c \teq 110 \meV$.
%This case is below the orange dashed line and the two quasiparticle bands overlaps.
%}
%\label{fig:ldos-off-2}
%\end{figure}
%% ------------------------------------------------------------------------------

We also checked cases with moderate $t_d/U$ but being farther away from the orange dashed
line: $t_d/U \teq 0.04,\ V^2/U \teq 0.8$ and $t_d/U \teq 0.08,\ V^2/U \teq 0.3$
(indicated by $\diamond$ and $\filleddiamond$ respectively in \Fref{fig:phase-diagram-pp}).
\Frefs{fig:ldos-td-0.04} (d) and (e) show the LDOS for the first case without and with
$\gamma_0$ included in the self-energy, and the LDOS for the latter case (without
and with $\gamma_0$ in the self-energy) are presented in \Frefs{fig:ldos-td-0.08}(d) and (e).
The first case is above the orange dashed line in \Fref{fig:phase-diagram-pp} with a large $J_K$,
and the two quasiparticle bands are separated in energy.
So the LDOS (\Fref{fig:ldos-td-0.04}(d)) has a gap sandwiched by two peaks.
In the later case, which is below the orange dashed line, the two quasiparticle bands overlap
with each other and there is a flat peak in LDOS (see \Fref{fig:ldos-td-0.08}(d)).
Once $\gamma_0$ is introduced for both cases, the LDOS changes into a single peak behaviour for 
both cases (\Fref{fig:ldos-td-0.04}(e) and \Fref{fig:ldos-td-0.08}(e)).
The fitting of LDOS width to the experimental data for these two cases are shown
in \Fref{fig:ldos-td-0.04}(f) and \Fref{fig:ldos-td-0.08}(f).
One can see that while the parameter $a$ for $t_d/U \teq 0.04$ still gives a reasonable fit,
the value of $a$ for $t_d/U \teq 0.08,\ V^2/U \teq 0.3$ is too small and the width
cannot be well fitted by \Eqref{eq:width-fit}.
We conclude that as $t_d/U$ increases, the fit deteriorates, especially  away from the orange
dashed line.

\begin{figure*}
\centering
\includegraphics[width=0.98\textwidth]{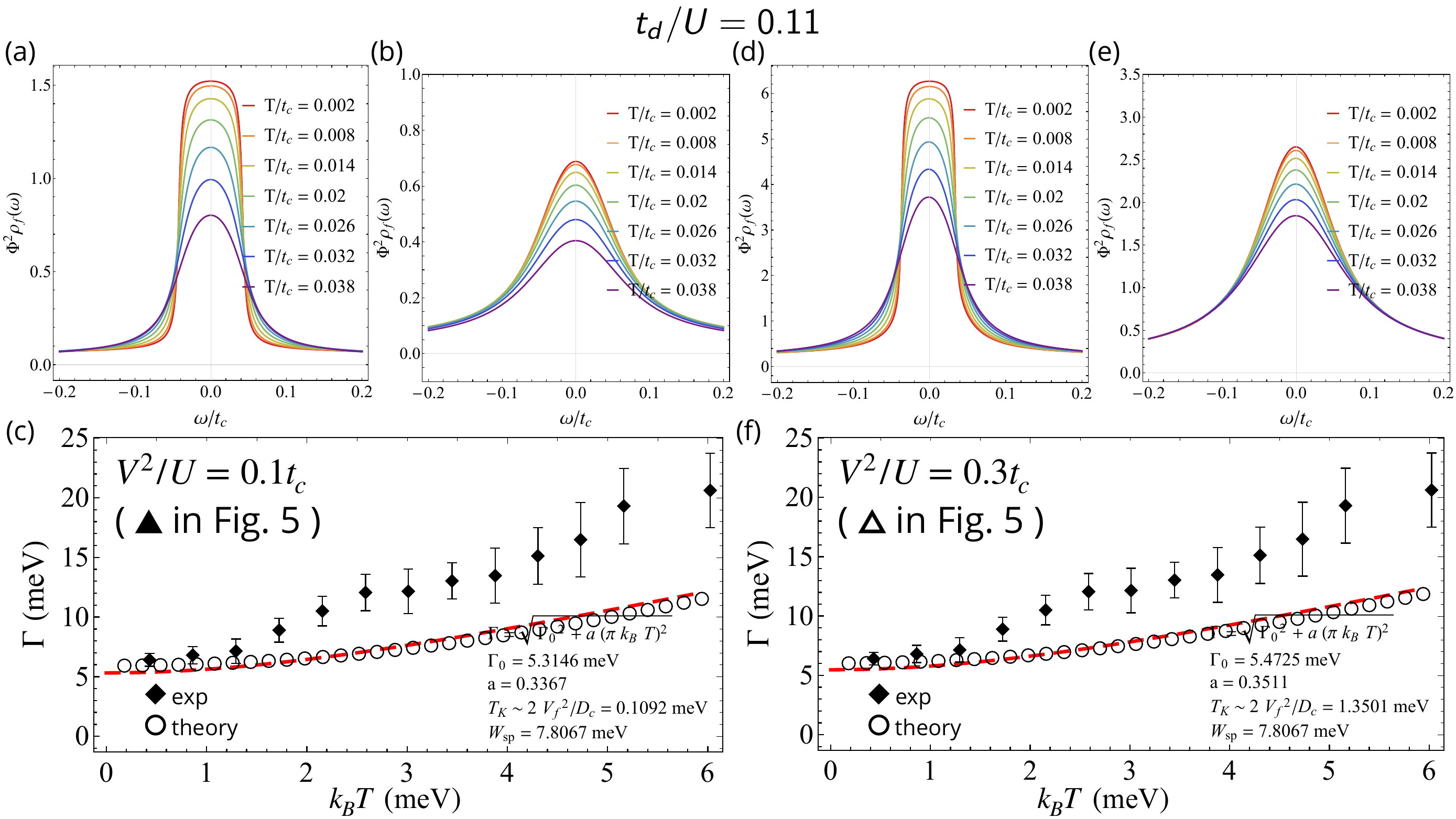}
\caption{
LDOS at $t_d/U \teq 0.11$.
(a)-(c) With $V^2/U \teq 0.1\, t_c$ (indicated by $\blacktriangle$ in
\Fref{fig:phase-diagram-pp}).
(a)-(b) LDOS without/with impurity scattering in the self-energy.
(c) Fitting of the width to experiment with $t_c \teq 90 \meV$.
The slope of the theoretical data is too small to fit the
experimental data.
(d)-(f) With $V^2/U \teq 0.3\, t_c$ (indicated by $\bigtriangleup$ in
\Fref{fig:phase-diagram-pp}).
(d)-(e) LDOS without/with impurity scattering in the self-energy.
(f) Fitting of the width to experiment with with $t_c \teq 90 \meV$.
Similar to the previous case, the slope of the theoretical data is too small to fit the
experimental data.
}
\label{fig:ldos-td-0.11}
\end{figure*}

Finally, for large $t_d/U$ (here we take $t_d/U \teq 0.11$) close to the critical value for the
metal-insulator transition in the isolated Hubbard model,
the LDOS for $V^2/U = 0.1\, t_c$ and $V^2/U = 0.3\, t_c$
(indicated by $\blacktriangle$ and
$\bigtriangleup$ separately in \Fref{fig:phase-diagram-pp}) are shown in
\Fref{fig:ldos-td-0.11}(a)-(c) and (d)-(f).
As expected, the LDOS has a flat top near $\omega \tsim 0$ without the inclusion of
$\gamma_0$ in the self-energy (\Fref{fig:ldos-td-0.11}(a) and \Fref{fig:ldos-td-0.11}(d)),
and will be broadened once $\gamma_0$ is introduced
(\Fref{fig:ldos-td-0.11}(b) and \Fref{fig:ldos-td-0.11}(e)).
\Fref{fig:ldos-td-0.11}(c) and \Fref{fig:ldos-td-0.11}(f) show the width for these cases and we see
that the experimental
data cannot be fitted by the theoretical results in this regime because the theoretical slope is
too small.

To summarize, by including a Fermi liquid type of (imaginary) self-energy into heavy
quasiparticles' Green function, it is possible to obtain a single-peak behaviour for the LDOS
even in the Anderson limit. By modifying the value of $\gamma_0$, the width of LDOS can be
well fitted by \Eqref{eq:width-fit}, which is the formula for a single impurity Kondo problem, as illustrated in \Fref{fig:ldos-anderson}(d).
Moreover, adjusting $t_c$ to fit the experimental width value at the lowest temperature,
our theory suggests that the experimental situation may be in or close to the
Anderson limit of the model. On the other hand, for intermediate $t_d/U$ a reasonable fit can be
obtained when the Kondo scale $J_K$ and the Heisenberg scale $J_H$ compete, resulting in a low 
temperature width which is smaller than $J_K$ or $J_H$,
as illustrated in \Fref{fig:ldos-td-0.04}(c).
In addition, our theory predicts
$a \tsim 0.3$ if the
hopping of the $d$ electrons is close to the critical value of for the metal-insulator transition
in isolated Hubbard model, a value which does not fit the experimental data.
% ------------------------------------------------------------------------------
% FIGURE
% ------------------------------------------------------------------------------
% ------------------------------------------------------------------------------

%% ------------------------------------------------------------------------------
%% FIGURE
%% ------------------------------------------------------------------------------
%\begin{figure}
%\centering
%\includegraphics[width=0.5\textwidth]{Fig-16}
%\caption{
%LDOS at $t_d/U \teq 0.11,\ V^2/U \teq 0.3\, t_c$ (indicated by $\bigtriangleup$ in
%\Fref{fig:phase-diagram-pp}).
%(a)-(b) LDOS without/with impurity scattering in the self-energy.
%(c) Fitting of the width to experiment with with $t_c \teq 90 \meV$.
%Similar to the previous case, the slope of the theoretical data is too small to fit the
%experimental data.
%}
%\label{fig:ldos-mit-2}
%\end{figure}
%% ------------------------------------------------------------------------------

\section{Summary and Discussions}\label{sec:discuss}
We have studied a model of coupled correlated and itinerant electrons which naturally
interpolates between the periodic Anderson model and the Hubbard model.
Using a slave rotor mean-field approach we have obtained a phase diagram that summarizes
the competition between a spinon Fermi surface state weakly coupled to a metal and an
interlayer coherent heavy Fermi liquid metallic state (illustrated in
\Frefs{fig:phase-diagram-pp}, \ref{fig:pb-chi-ph} and \ref{fig:phi-ph-xi-1}).
In the localized or atomic limit where our model reduces to the periodic Anderson model,
the Kondo coupling needed to destroy the spin liquid in favor of the metal,
$J_K \tsim V^2/U$, has a logarithmic dependence on the hopping of the correlated electrons
in the putative spin liquid layer $t_d/U$, reflecting that the emergent scales determining the
competition are the Kondo temperature $T_K \tsim \rho^{-1} e^{-1/J_K\rho}$ ($\rho \tsim t_c^{-1}$) and Heisenberg coupling
$J_H \tsim t_d^2/U$.
Therefore, although technically in such limit the spin liquid is destabilized via a weak coupling
instability, the critical Kondo coupling needed to destabilize the spin liquid grows rather
fast with the Heisenberg coupling, giving rise to the rapid rise of the boundary between the
spin liquid and the heavy metal at small $t_d/U$ seen in \Frefs{fig:phase-diagram-pp},
\ref{fig:pb-chi-ph} and \ref{fig:phi-ph-xi-1}. In this limit one can use the measured saturation
width $T_K$ to place an upper bound on the Heisenberg coupling $J_H$, resulting in a rather
small bound of about $5 \meV$ from the experiments of Ref.~\cite{Wei2020}.
On the other hand, at larger values of $t_d/U \tsim 0.1$ when the spin liquid has a sizable
bandwidth, the critical $J_K$ is comparable to $t_d/U$,
and near the Mott transition the critical Kondo coupling needed to destabilize the spin liquid
vanishes linearly with the distance of $t_d/U$ away from the critical value associated with the
Mott  metal-insulator-transition, at mean field level. However, we find that generically the peak
width is dominated by the spinon bandwidth, leading to a width that is too broad and with too 
weak a temperature dependence to explain the data. The exception is when the system happens 
to fall near the crossover line indicated in orange in \Fref{fig:phase-diagram-pp}, where a 
reasonable fit to the data can also be obtained. In this case, the Kondo scale $J_K$ and the 
Heisenberg scale $J_H$ compete, giving rise to a narrow peak with a width which is smaller than 
either scale at low temperature. As a result, in this case the low temperature width cannot be used 
as a bound for either scale, and  it is possible that $J_H$ is much larger than the $5 \meV$ bound
mentioned previously.

The above conclusion was reached by studying the LDOS of the heavy metal throughout this
phase diagram, which can be directly accessed via STM experiments \cite{Wei2020}.
In the local moment periodic Anderson limit of the model the coherent hybridization of
correlated and itinerant electrons in the heavy metal leads to
the bare LDOS acquiring a two-peak structure due to the opening of a direct optical band gap.
On the other hand, near the Mott-metal-insulator transition the LDOS
features a rather flat shape due to a relatively large spinon band width.
The measured LDOS is however further broadened by the intrinsic lifetime of the heavy
quasi-particles arising from their interactions and also by disorder,
leading to a smearing of the double-peak structure in the periodic Anderson model limit.
We have argued that including these effects renders the double peak structure effectively
into a single peak, and we have found good agreement with the shape and temperature
dependence of the peak reported in recent STM experiments \cite{Wei2020}, as illustrated in \Fref{fig:ldos-anderson}(d).
We also find reasonable fit to the data at intermediate $t_d/U$ in the vicinity of the orange line
in \Fref{fig:phase-diagram-pp}, as illustrated in \Fref{fig:ldos-td-0.04}(c).
%lending support to the idea that the system comprised of a putative \TaSe\ spin liquid residing
%on top of the metallic 1H-TaSe\textsubscript{2} realizes an interlayer coherent heavy metal
%that is closer to the localized periodic Anderson model limit.

We note that in the localized limit of small $t_d/U$ the Hubbard model in the triangular lattice
is expected not to form
a spinon Fermi surface state, but to order into a conventional $120^\circ$ AFM phase. This piece
of physics is not captured in our slave rotor mean-field theory, which favors spin disordered
ground states.
Therefore, our results pose a challenge for the interpretation of the behavior of the stand-alone 
putative \TaSe\ as a quantum spin liquid: if indeed the system is near the Anderson limit, this
raises the possibility that it could be instead comprised of
localized moments that are rather weakly coupled and might ultimately weakly
order at yet lower temperatures in cleaner samples.
We however caution that we cannot definitely rule out that the putative spin liquid layer is at an 
intermediate coupling strength $t_d/U$ that brings the system closer
to the Mott transition, where also a small interlayer tunnelling can destabilize the spin liquid. An 
additional consideration is that the actual \TaSe\ system involves multiple bands and is probably 
not described by a single band Mott-Hubbard model. While the spin liquid is stabilized only near 
the Mott transition in a single band model \cite{Lee2005}, it is possible that a multi-band 
description can extend the spin liquid to lower effective $t_d$.

Additionally, to reiterate the potential uncertainties, we wish to note that the parameter $a$ in 
\Eqref{eq:width-fit} that we used near the Mott transition has a Fermi liquid form but it
can be changed by tuning the value of $\gamma_0$ and $E_0$,
which are respectively controlled by disorder and quansiparticle interactions,
and hence are inherently difficult scales to estimate accurately.

We want also to point out that in our calculation, we considered the metallic electrons to have
the same lattice constant and Brillouin zone as the correlated electrons.
In doing so, we are imagining that in a more microscopic description one would be folding
the Brillouin of the metallic 1H-TaSe\textsubscript{2}, which does match with the smaller 
Brillouin zone of the star of David structure of \TaSe, and that after this one is only including
one of the folded bands of itinerant electrons.
However, the hybridization with electrons at higher energy scales (coming from other
folded bands) could also play an important role in determining the phase boundary and
the form of LDOS, but such details lie beyond the scope of the considerations that we have
explored in the present work.

% Finally, we want to emphasize that although the Hubbard-Anderson
% model in this paper is introduced for the study of coupled TaSe\textsubscript{2} layers, it also describes the physics of some other
% realistic systems, e.g., PdCoO\textsubscript{2} \cite{Sunko2020}.
%

\begin{acknowledgments}
We thank Michael F. Crommie, Wei Ruan and Yi Chen for sharing their data and discussions.
We also thank Peng Rao for fruitful discussions. PAL acknowledges support by
DOE office of Basic Sciences grant number DE-FG02-03ER46076.
\end{acknowledgments}

\appendix

\section{Finite Temperature Rotor Mean Field Approach}\label{sec:append-rotor}
As mentioned in the main text, for the order parameter of metallic phase,
$\Phi \teq \langle e^{i\theta} \rangle$, we estimate its value by
taking the average with respect to a single site Hamiltonian:
\begin{align}
H_{\theta}^{(1)} &= -K_\theta \left( e^{i\theta} + e^{ -i\theta} \right)
+ \frac{U}{2} n_{\theta}^2 \\
&= H_K + H_U,
\end{align}
where $H_K = -K_\theta \left( e^{i\theta} + e^{ -i\theta} \right)$
and $H_U = \frac{U}{2} n_{\theta}^2$. We have taken $\lambda \teq 0$
to fulfil the constrain \Eqref{eq:constrain} and the half-filling of the spinon.
Because we are interested in the large $U$ limit of the model ($t_d/U \tlapprox 1/8$),
it is reasonable to use a first-order perturbation (in $H_K$) to estimate the expectation value:
\begin{align}
\langle e^{i\theta} \rangle &= \frac{\Tr \left( e^{-\beta (H_U+H_K)} e^{i\theta} \right)}{\Tr \left(
e^{-\beta(H_U+H_K)} \right)} \nonumber \\
& \approx -\int_0^\beta d\tau \Tr \left( e^{-\beta H_U} e^{\tau H_U} H_K e^{-\tau H_U} e^{i\theta}
\right) / \Tr \left( e^{-\beta H_U} \right),
\end{align}
one can take the trace with the eigenbasis of angular momentum $n_{\theta}$:
 $\left\lbrace | n \rangle \right\rbrace$, which satisfies:
 $n_{\theta} | m \rangle \teq m | m \rangle$ and
 $e^{i\theta}| n \rangle \teq | n+1 \rangle$, and we will denote the eigenvalue of
 $H_U$ by $E_n \teq \frac{U}{2} n^2$.
 It is straightforward to obtain:
 \begin{subequations}
 \begin{align}
& -\int_0^\beta d\tau \Tr \left( e^{-\beta H_U} e^{\tau H_U} H_K e^{-\tau H_U} e^{i\theta}
\right) \nonumber \\
& = K_\theta \sum_n \int_0^\beta d\tau e^{-\beta E_n} e^{-\beta E_n} e^{\tau (E_n-E_{n+1})}
\nonumber \\
& = K_\theta \sum_n \frac{e^{-\beta E_{n+1}} - e^{-\beta E_n}}{E_n - E_{n+1}},
 \end{align}
 \begin{equation}
 \Tr \left( e^{-\beta H_U} \right) = \sum_n e^{-\beta E_n},
 \end{equation}
 \end{subequations}
so one finally arrives at:
\begin{align} \label{eq:phi-K}
\langle e^{i\theta} \rangle &\approx \chi_{\theta,1} K_\theta, \\
\chi_{\theta,1} &= \sum_n \frac{e^{-\beta E_{n+1}} - e^{-\beta E_n}}{E_n - E_{n+1}}/
\sum_n e^{-\beta E_n}.
\end{align}
By Taking the zero temperature limit, one can recover the zero temperature result given by:
\begin{equation}
\lim_{\beta \rightarrow \infty} \chi_{\theta,1}(\beta) \teq 4/U.
\end{equation}
Next, we extrapolate the expression above, which is valid only for small $K_\theta$,
with the phenomenological formula:
\begin{equation}
\langle e^{i\theta} \rangle = \frac{K_\theta}{\sqrt{ \chi_{\theta,1}^{-2} + K_\theta^2 }},
\end{equation}
which 
recovers the behavior from \Eqref{eq:phi-K} at small $K_\theta$ and also the approach of 
$\langle e^{i\theta} \rangle \rightarrow 1$, which is expected at large $K_\theta$
(and it is also consistent with the constraint that $\langle e^{i\theta} \rangle \leq 1$).

For $\langle e^{-i\theta_i} e^{i\theta_j} \rangle$, one can perform same kind of calculation.
We estimate it by taking the expectation value with respect to the Hamiltonian:
\begin{equation}
\tilde{H}_\theta = \frac{U}{2}\sum_i n_{\theta,i}^2 - 2 T_\theta \sum_{\langle i,j \rangle}
\left( e^{-i\theta_i} e^{i\theta_j} + h.c. \right),
\end{equation}
taking $T_\theta$-term as a perturbation, after some algebra, one obtains that:
\begin{align} \label{eq:phi-phi}
& \langle e^{-i\theta_i} e^{i\theta_j} \rangle \approx \chi_{\theta,2} T_\theta, \\
& \chi_{\theta,2} = 2\left( \sum_{n_i \neq n_j+1} \frac{e^{-\beta(E_{n_i-1}+E_{n_j+1})}
- e^{-\beta(E_{n_i}+E_{n_j} ) }}{E_{n_i}+E_{n_j} - (E_{n_i-1} + E_{n_j+1}) } \right. \nonumber \\
& \quad \quad \left. + \sum_n \beta e^{ -\beta(E_n+E_{n-1}) } \right)/
\left( \sum_n e^{-\beta E_n} \right)^2,
\end{align}
and for the zero temperature limit, one recovers the value $\chi_{\theta,2} = 4/U$.
Because we are interested in small $t_d$ limit (remember that $T_\theta = t_d \chi_0$), we simply
use \Eqref{eq:phi-phi} throughout our calculations.

It should be noted that the current mean-field would 
predicts an artificial second order phase transition for
any low temperature phase with finite
$\langle e^{i \theta} \rangle$ to a high temperature
phase with $\langle e^{i \theta} \rangle \teq 0$,
similar to the case of slave boson descriptions at 
mean-field level~\cite{Coleman1984}.
In reality, there is no such phase
transition as a function of temperature but only a
crossover~\cite{Coleman1985,Franco2002},
and the expectation value of
$\langle e^{i \theta} \rangle$
is always finite at non-zero temperatures.
However, the zero temperature 
transitions which are the focus of the 
main manuscript are allowed to be
sharp second order phase transitions in
principle~\cite{Senthil2004,Senthil2008}.

\section{Tunnelling DOS of the single impurity Anderson Model}\label{sec:append-ldos}
In this section, we briefly review the theory of tunnelling DOS for a single impurity
Anderson model and give a more thorough deriviation on the fitting of STM results expanding on  the previous studies by Nagaoka
et al. \cite{Nagaoka2002}.

For a single impurity Anderson model, one can calculate the tunnelling DOS of the local
electron using perturbation theory since there is \emph{no} phase transition as
the on-site interaction $U$ increases \cite{Coleman2015}.
Early theoretical calculations \cite{Yamada1975,Costi1994} showed that the (retarded)
Green function of the local electron for the particle-hole symmetric case reads
(valid at small $\omega$ and $T$):
\begin{align} \label{eq:rho_d2}
G_d(\omega, T) & = \frac{1}{\omega - \epsilon_d - \Re\Sigma(\omega) + i \Delta
- i \Im \Sigma(\omega, T)} \nonumber \\
& = \frac{Z}{\omega - \tilde{\epsilon}_d + i Z ( \Delta - \Im \Sigma(\omega, T) )},
\end{align}
where 
\begin{subequations}
\begin{align}
%\begin{subequations}
\tilde{\epsilon}_d & = \epsilon_d + \Re \Sigma( 0 ) \approx 0, \\
\Im \Sigma( \omega, T ) & = -\frac{\Delta}{2} \alpha^2
\left[ \left( \frac{\omega}{T_K} \right)^2 + \pi^2 \left( \frac{T}{T_K} \right)^2 \right], 
%\alpha & = \pi/4.
%\end{subequations}
\end{align}
\end{subequations}
where $\alpha$ is a number of order unity and equals $\pi/4$.
%in mean field theory.
In the second line of \Eqref{eq:rho_d2} we follow standard practice and expand $\Re\Sigma(\omega)$  to linear order in $\omega$ near the pole
with
\begin{equation}
Z  = \frac{1}{1 - \partial_\omega \Re \Sigma(\omega)|_{\omega = 0}} = \frac{T_K}{\alpha \Delta}, \\
\end{equation}
Then it is straightforward to obtain the spectral 
function:
\begin{align} \label{eq:rho_d}
\rho_d (\omega) & = \frac{Z^2}{\pi} \frac{\left( \Delta - \Im \Sigma(\omega) \right)}
{\omega^2 + Z^2 \left( \Delta - \Im \Sigma(\omega) \right)^2} \nonumber \\
& = \frac{Z^2 \Delta}{\pi}
\frac{1+\frac{1}{2} \alpha^2
\left( \frac{\omega^2}{T_K^2} +  \frac{\pi^2 T^2}{T_K^2} \right) }{
\omega^2 + Z^2 \Delta^2 \left( 1 + \frac{1}{2} \alpha^2
\left( \frac{\omega^2}{T_K^2} +  \frac{\pi^2 T^2}{T_K^2} \right) \right)^2 } \nonumber \\
& = \frac{1}{\pi \Delta}
\frac{1}{
\frac{ \omega^2/(Z \Delta)^2 }{1+\frac{1}{2} \alpha^2
\left( \frac{\omega^2}{T_K^2} +  \frac{\pi^2 T^2}{T_K^2} \right)} + 1 + \frac{1}{2} \alpha^2
\left( \frac{\omega^2}{T_K^2} +  \frac{\pi^2 T^2}{T_K^2} \right) } \nonumber \\
& = \frac{1}{\pi \Delta}
\frac{1}{
\frac{ \alpha^2 \omega^2/T_K^2}{1+\frac{1}{2} \alpha^2
\left( \frac{\omega^2}{T_K^2} +  \frac{\pi^2 T^2}{T_K^2} \right)} + 1 + \frac{1}{2} \alpha^2
\left( \frac{\omega^2}{T_K^2} +  \frac{\pi^2 T^2}{T_K^2} \right) }.
\end{align}
In the previous work by Nagaoka et al.\cite{Nagaoka2002}, they did not include the expansion near the pole, which amounts to setting $Z \teq 1$. With this and %improperly 
setting
$\alpha \teq 1$, they argued that the $\omega$ term in the denominator of \Eqref{eq:rho_d2} can be dropped and they arrive at the incorrect result that $\rho_d \propto 1/\Im \Sigma(\omega, T)$, i.e.:
\begin{equation} \label{eq:nagaoka-rho_d}
\rho_d(\omega) = \frac{1}{\pi \Delta}
\frac{1}{  
 1 + \frac{1}{2} \left( \frac{\omega^2}{T_K^2} +  \frac{\pi^2 T^2}{T_K^2} \right)
}
\end{equation}
with the prediction that the width reads:
\begin{equation} \label{eq:width-exp}
\Gamma_{exp} = \sqrt{2T_K^2 + \pi^2 T^2},
\end{equation}
which suggests the slope of $\Gamma$ with respect to $T$ is approximately $\pi$ for $T \gg T_K$

\begin{figure}
\centering
\includegraphics[width=0.46\textwidth]{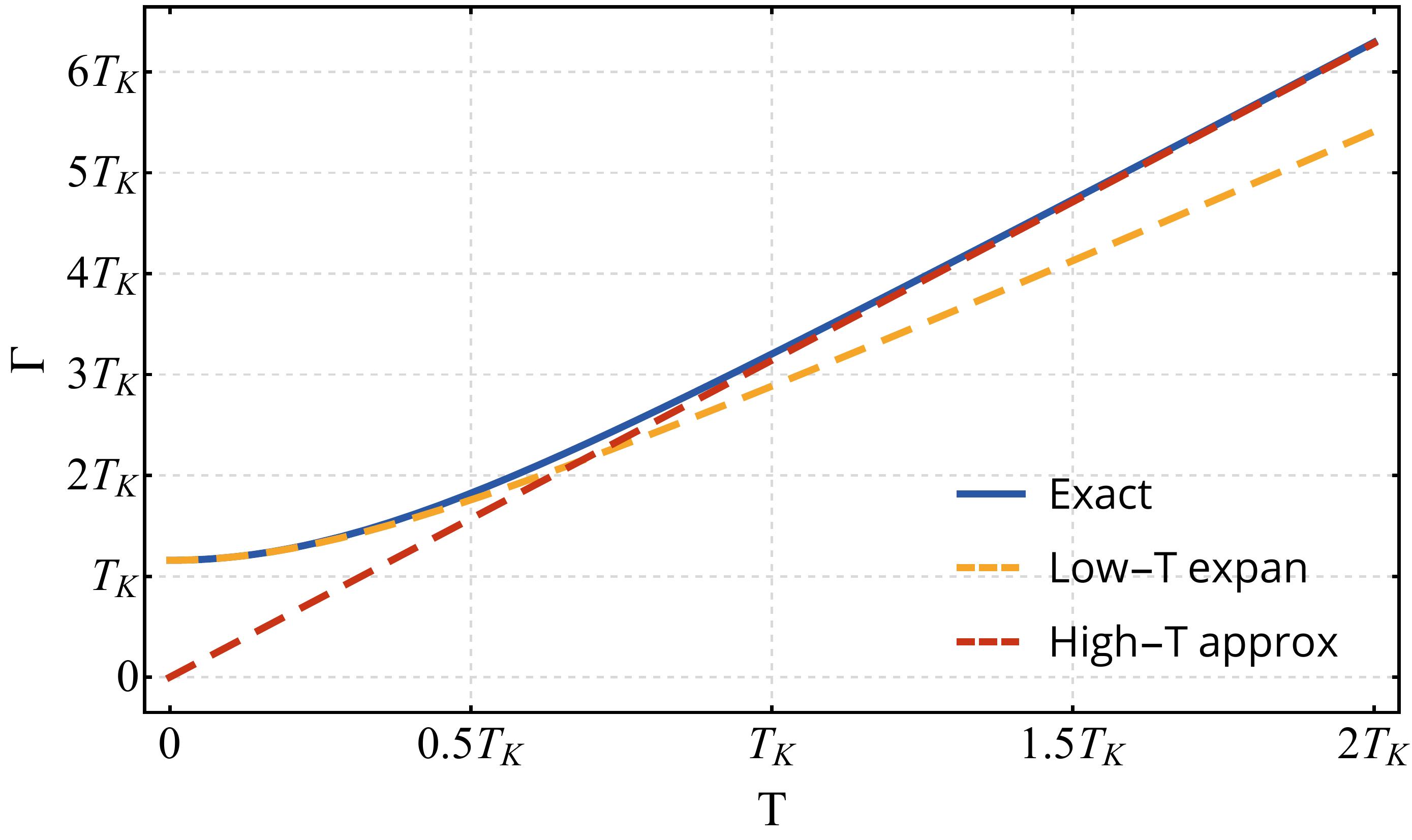}
\caption{
Plot of the LDOS width $\Gamma$ with respect to the temperature $T$ based on
Fermi liquid theory.
The blue solid line stands for the exact result \Eqref{eq:Gamma exact},
the orange dashed line indicates the low temperature expanded form \Eqref{eq:Gamma-low T}
and the red dashed line shows the high temperature approximated form
\Eqref{eq:Gamma-high T}.
}
\label{fig:width-exact}
\end{figure}

However, as we can see from the second line in \Eqref{eq:rho_d2}, due to the fact that  $Z \approx T_K/\Delta \ll 1$, $\omega$ cannot be dropped.
This is seen explicitly in \Eqref{eq:rho_d}, where the 
first term in the denominator dropped by Nagaoka
et al.\cite{Nagaoka2002} is clearly of the
same order as the rest and should be kept. Nevertheless, we shall show below that conclusion that the slope of $\Gamma$ with respect to $T$ is approximately $\pi$ at relatively
high temperature is actually valid. The more complete \Eqref{eq:rho_d} implies that the lineshape is not a simple Lorentzian. Instead, we  calculate the half-width at half height
%Since the $\rho_d(\omega)$ is a decreasing function as $|\omega|$ increases, the width
by requiring
$\Gamma$ to satisfy $\rho_d(\Gamma) = \rho_d(0)/2$, which leads to:
\begin{equation}
\alpha^2 \left( \frac{\Gamma}{T_K} \right)^2
= \left( 1 + \frac{1}{2}\alpha^2 \left( \frac{\pi T}{T_K} \right)^2 \right)^2
- \left( \frac{1}{2} \alpha^2 \left( \frac{\Gamma}{T_K} \right)^2 \right)^2,
\end{equation}
after some algebra, one can show that
\begin{equation} \label{eq:Gamma exact}
\Gamma = \frac{\sqrt{2}}{\alpha}
\left( \sqrt{T_K^4 + \left( T_K^2 + \frac{1}{2}\alpha^2 \pi^2 T^2\right)^2 } - T_K^2 \right)^{1/2}.
\end{equation}
In the low temperature limit, the width can be approximated as
\begin{equation} \label{eq:Gamma-low T}
\Gamma \approx \sqrt{ \frac{2(\sqrt{2}-1)}{\alpha^2}T_K^2 + \frac{1}{\sqrt{2}} \pi^2 T^2},
\end{equation} 
on the other hand, for large $T$ such that $T \gg T_K$,
$\Gamma$ can be approximated as
\begin{equation} \label{eq:Gamma-high T}
\Gamma \approx \pi T.
\end{equation}
Going back to \Eqref{eq:rho_d}, we see that in large $T$ limit the first term in the denominator becomes a \textit{nonzero} constant. It affects the effective definition of the zero temperature width in terms of $T_K$, but does not affect the high temperature limit of the line-width.
Although the low temperature expansion \Eqref{eq:Gamma-low T} seems to suggest
that the slope of  $\Gamma$-$T$ curve would saturate to $\pi/2^{1/4}$ at relatively large 
temperatures (see the orange dashed line in \Fref{fig:width-exact}),
the slope derived from \Eqref{eq:Gamma exact} actually saturates to $\pi$ at
higher temperatures, as indicated by the blue line in \Fref{fig:width-exact}.

% ------------------------------------------------------------------------------
% FIGURE
% ------------------------------------------------------------------------------
% ------------------------------------------------------------------------------

According to \Eqref{eq:Gamma exact}, the zero temperature width should be
$\Gamma(T \teq 0) \teq [2 ( \sqrt{2} - 1) ]^{\frac{1}{2}}/\alpha \approx 1.16 T_K$, while
\Eqref{eq:width-exp} predicts $\Gamma_{exp}(T \teq 0) \teq \sqrt{2}T_K$. Therefore, with a given
set of experimental data of $\Gamma$ versus $T$, the extracted $T_K$ using
\Eqref{eq:width-exp} would be slightly smaller than the one predicted from
\Eqref{eq:Gamma exact}.
On the other hand, both expressions suggest that $\Gamma$ has an approximately linear
dependence on $T$ for $T \tsim T_K$ with a $\pi$ slope.

Finally, comparing the Fermi liquid theory presented above and the more exact numerical
renormalization group (NRG) calculation, one can see
that both theories imply that the LDOS is not
a simple Lorentzian form. The fermi liquid theory
suggests $\Gamma(T \teq 0) \approx 1.16 T_K$ while
the NRG suggests $\Gamma(T \teq 0) = T_K$.
The NRG LDOS curve can be quantitatively well fitted by a phenomenological
expression suggested by Frota and Oliveira \cite{Frota1986,Frota1992}:
\begin{align}
\rho_f(\omega) & = \frac{2}{\pi \Gamma_A} \text{Re}\left[ 
\left( \frac{\omega + i \Gamma_K}{i \Gamma_K} \right)^{-1/2}
\right] \nonumber \\
& = \frac{2}{\pi \Gamma_A} \left(
\frac{1+\sqrt{1+(\omega/\Gamma_K)^2}}{2(1+(\omega/\Gamma_K)^2)}
\right)^{1/2},
\end{align}
with $\Gamma_A$ and $\Gamma_K$ being fitting parameters.
However, it should be noted that this formula
is a phenomenological parameterization of the model
and is not able to predict the temperature
dependence of LDOS and its width \cite{Frota1992}.

\bibliographystyle{apsrev4-2}
\bibliography{Hubbard-metal}

\end{document}